\documentclass[graybox]{svmult}

\usepackage{mathptmx}       
\usepackage{helvet}             
\usepackage{courier}            
\usepackage{type1cm}         
\usepackage{makeidx}         
\usepackage{graphicx}         
\usepackage{multicol}          
\usepackage[bottom]{footmisc}
\usepackage{amssymb}
\usepackage{bm}

\makeindex             

\begin{document}

\title*{Magnetic Catalysis: A Review}

\author{Igor A. Shovkovy}
\institute{Igor A. Shovkovy 
\at Department of Applied Sciences and Mathematics,
Arizona State University, Mesa, AZ 85212, 
\email{Igor.Shovkovy@asu.edu}}
%
%

\maketitle

\abstract*{We give an overview of the magnetic catalysis phenomenon. In the framework 
of quantum field theory, magnetic catalysis is broadly defined as an enhancement 
of dynamical symmetry breaking by an external magnetic field. We start from a 
brief discussion of spontaneous symmetry breaking and the role of a magnetic 
field in its a dynamics. This is followed by a detailed presentation of the essential 
features of the phenomenon. In particular, we emphasize that the dimensional 
reduction plays a profound role in the pairing dynamics in a magnetic field. Using 
the general nature of underlying physics and its robustness with respect to 
interaction types and model content, we argue that magnetic catalysis is a 
universal and model-independent phenomenon. In support of this claim, we 
show how magnetic catalysis is realized in various models with short-range 
and long-range interactions. We argue that the general nature of the 
phenomenon implies a wide range of potential applications: from certain 
types of solid state systems to models in cosmology, particle and nuclear 
physics. We finish the review with general remarks about magnetic 
catalysis and an outlook for future research.}

\abstract{We give an overview of the magnetic catalysis phenomenon. In the framework 
of quantum field theory, magnetic catalysis is broadly defined as an enhancement 
of dynamical symmetry breaking by an external magnetic field. We start from a 
brief discussion of spontaneous symmetry breaking and the role of a magnetic 
field in its a dynamics. This is followed by a detailed presentation of the essential 
features of the phenomenon. In particular, we emphasize that the dimensional 
reduction plays a profound role in the pairing dynamics in a magnetic field. Using 
the general nature of underlying physics and its robustness with respect to 
interaction types and model content, we argue that magnetic catalysis is a 
universal and model-independent phenomenon. In support of this claim, we 
show how magnetic catalysis is realized in various models with short-range 
and long-range interactions. We argue that the general nature of the 
phenomenon implies a wide range of potential applications: from certain 
types of solid state systems to models in cosmology, particle and nuclear 
physics. We finish the review with general remarks about magnetic 
catalysis and an outlook for future research.}


\section{Introduction}
\label{sec:1}

The magnetic catalysis is broadly defined as an enhancement of dynamical symmetry 
breaking by an external magnetic field. In this review, we discuss the underlying physics 
behind magnetic catalysis and some of its most prominent applications. Considering 
that the ideas of symmetry breaking take the center stage position in many branches of 
modern physics, we hope that this review will be of interest to a rather wide audience. 

In particle and nuclear physics, spontaneous symmetry breaking is commonly used 
in order to explain the dynamical origin of the mass of elementary particles. 
In this context, the idea was realized for the first time over 50 years ago  by Nambu 
and Jona-Lasinio \cite{Nambu:1961tp,Nambu:1961fr}, who suggested that 
``the nucleon mass arises largely as a self-energy of some primary 
fermion field through the same mechanism as the appearance of energy gap in the 
theory of superconductivity." As we now know, the analogy with superconductivity is 
not very close and the description of chiral symmetry breaking in terms of quarks may 
be more natural than in terms of nucleons. However, the essence of the dynamical mass 
generation was captured correctly in Refs.~\cite{Nambu:1961tp,Nambu:1961fr}. In fact, 
with the current state of knowledge, we attribute most of the mass of visible matter in 
the Universe to precisely this mechanism of mass generation, which is associated 
with breaking of the (approximate) chiral symmetry.  

The conceptual knowledge that the mass can have a dynamical origin opens myriads of 
theoretical possibilities that would appear meaningless in classical physics. For example, 
keeping in view the above mentioned mechanism of mass generation through chiral symmetry 
breaking, it is reasonable to suggest that the masses of certain particles can be modified or 
even tuned by proper adjustments of physical parameters and/or external conditions. 

One of the obvious knobs to control the value of the dynamical mass is an external 
magnetic field. In addition to be a good theoretical tool, magnetic fields are also 
relevant to many applications. For example, they are commonly present and 
play an important role in such physical systems as the Early Universe 
\cite{Vachaspati:1991nm,Enqvist:1993kf,Cheng:1994yr,Baym:1995fk,
Grasso:2000wj}, heavy ion collisions \cite{Kharzeev:2007jp,Skokov:2009qp}, 
neutron stars \cite{Thompson:1993hn,Duncan:1992hi}, 
and quasi-relativistic condensed matter systems like graphene 
\cite{Novoselov:2005kj,2005Natur.438..201Z}.

As we discuss in detail in this review, the magnetic field has a strong tendency to enhance 
(or ``catalyze") spin-zero fermion-antifermion condensates. Such condensates are commonly 
associated with breaking of global symmetries (e.g., such as the chiral symmetry in particle 
physics and the spin-valley symmetry in graphene) and lead to a dynamical
generation of masses (energy gaps) in the (quasi-)particle spectra. The corresponding 
mechanism is called {\em magnetic catalysis} \cite{Gusynin:1994re}. 

It should be emphasized that, in a striking contrast to its role in superconductivity, the 
magnetic field helps to {\em strengthen} the chiral condensate. There are many 
underlying reasons for its different role. Unlike the superconductors, the ground 
state with a nonzero chiral condensate 
shows no Meissner effect. This is because the chiral condensate can be thought of 
as a condensate of neutral fermion-antifermion pairs, not charged Cooper pairs that 
can give rise to supercurrents and perfect diamagnetism. Also, in a usual
Cooper pair, the two electrons  have opposite spins and, therefore, opposite magnetic 
moments. When placed in a magnetic field, only one of the magnetic moments can 
minimize its energy by orienting along the direction of the field. The other magnetic 
moment will be stuck in a frustrated position pointing in the opposite direction. 
This produces an energy stress and tends to break the Cooper pair. (Note, however, 
that the orbital motion plays a much more important role in breaking nonrelativistic Cooper 
pairs.) In a neutral spin-zero pair, in contrast, the magnetic moments of the fermion 
(with a fixed charge and spin) and the antifermion (with the opposite charge and spin) 
point in the same direction. Therefore, both magnetic moments can comfortably align 
along the direction of the magnetic field without producing any frustration in the pair. 
(Also, in relativistic systems the fermion-antifermion condensate is not destroyed 
by the orbital motion.)

The above explanation of the role that the magnetic field plays in strengthening the chiral 
condensate is semi-rigorous at best and does not capture all the subtleties of the dynamics
behind magnetic catalysis (e.g., completely leaving out the details of the orbital motion). It
does demonstrates, however, how the magnetic field can have, at least in principle, so drastically 
different effects on the dynamical generation of mass on the one hand and on superconductivity
on the other. (It may be curious to mention here that, in cold dense quark matter, it is possible 
to obtain color superconducting states, in which diquark Cooper pairs are neutral with respect 
to the in-medium (but not vacuum) electro-magnetism \cite{Gorbar:2000ms,Alford:1999pb}. 
In such quark matter, the in-medium magnetic field is not subject to the Meissner effect 
and, in fact, can enhance color superconductivity \cite{Noronha:2007wg,Fukushima:2007fc,
Ferrer:2006vw,Ferrer:2005pu,Manuel:2006gu,Fayazbakhsh:2010gc,Fayazbakhsh:2010bh,Yu:2012jn}.)

The early investigations of the effects of strong magnetic fields on chiral symmetry breaking 
in $(2+1)$- and $(3+1)$-dimensional models with local four-fermion interactions have 
appeared in late 1980s and early 1990s \cite{Klevansky:1989vi,Suganuma:1990nn,Klimenko:1990rh,
Klimenko:1991he,Schramm:1991ex,Krive:1992xh,Klevansky:1992qe}. In these studies, it was already 
found that a constant magnetic field stabilizes the chirally broken vacuum state.  

The explanation of the underlying physics was given in Ref.~\cite{Gusynin:1994re}, where 
the essential role of the dimensional reduction, $D\to D-2$, in the low-energy dynamics 
of pairing fermions in a magnetic field was revealed. As a corollary, it was also established
that the presence of a magnetic field leads to the generation of a dynamical mass even at the weakest 
attractive interaction between fermions \cite{Gusynin:1994re,Gusynin:1994xp,Gusynin:1995gt,
Gusynin:1994va}. The general nature of the underlying physics was so compelling 
that it was suggested that the corresponding dynamical generation of the chiral condensate 
and the associated spontaneous symmetry breaking in a magnetic field are universal and 
model-independent phenomena. To emphasize this fact, the new term ``magnetic catalysis" 
was coined \cite{Gusynin:1994re}. 

The model-independent nature of magnetic catalysis was tested in numerous $(2+1)$- 
and $(3+1)$-dimensional models with local four-fermion interactions \cite{Babansky:1997zh,
Klimenko:1998su,Ebert:1999ht,Vdovichenko:2000sa,Zhukovsky:2000yd,Inagaki:2003yi,
Inagaki:2004ih,Ghosh:2005rf,Osipov:2007je,Hiller:2008eh,Klimenko:2008mg,Menezes:2008qt,
Menezes:2009uc,Fayazbakhsh:2010bh,Chatterjee:2011yi,Avancini:2012ee,Ferrari:2012yw},
including models with additional gauge interactions \cite{Ishii:2001xg}, 
higher dimensional models \cite{Gorbar:2000ku},
${\cal N}=1$ supersymmetric models \cite{Elias:1996zu}, 
quark-meson models \cite{Andersen:2011ip,Andersen:2012bq}, 
models in curved space \cite{Gitman:1996mk,Geyer:1996np,Inagaki:1997nv} 
and QED-like gauge theories \cite{Gusynin:1995gt,Leung:1995mh,
Parwani:1994an,Parwani:1995am,Gusynin:1997kj,Gusynin:1998zq,Farakos:1998ds,Lee:1998sr,
Gusynin:1999pq,Farakos:1999qc,Gusynin:1998nh,Gusynin:2000tv,Alexandre:2000yf,Alexandre:2001vu,
Gusynin:2003dz,Sadooghi:2007ys,Ayala:2009fv,Ayala:2010fm}. The realization of magnetic catalysis was 
investigated in chiral pertubation theory \cite{Shushpanov:1997sf,Cohen:2007bt,Cohen:2008bk} and in QCD 
\cite{Miransky:2002rp,Kabat:2002er}, as well as in a models with the Yukawa interaction 
\cite{Ferrer:1998fd,Ferrer:1998xp,Ferrer:2000ed,Elizalde:2002ca}. There are studies of  
magnetic catalysis using the methods of the renormalization group 
\cite{Semenoff:1999xv,Scherer:2012nn,Fukushima:2012xw}, 
lattice calculations \cite{Farakos:1998fe,Buividovich:2010tn,
Cea:2011hu,2010PhRvD..82e1501D,2011PhRvD..83k4028D,Bali:2011qj,Bali:2011uf,Bali:2012zg}
and holographic dual modelss of large-$N$ gauge theories 
\cite{Zayakin:2008cy,Filev:2009xp,Filev:2010pm,Preis:2010cq,Evans:2010hi,Evans:2010iy,
Evans:2010xs,Evans:2011mu,Evans:2011tk,Preis:2011sp,Erdmenger:2011bw,
Filev:2011mt,Bolognesi:2011un,Bolognesi:2012pi,Alam:2012fw}. 
Similar ideas were extended to solid state systems describing high-temperature supersonductivity  
\cite{Liu:1998mg,Semenoff:1998bk,Zhukovsky:2000jm,Ferrer:2001fz,Ferrer:2002gf,Zhukovsky:2003bs},
highly oriented pyrolitic graphite \cite{Khveshchenko:2001zza,Khveshchenko:2001zz,Gorbar:2002iw},
as well as monolayer \cite{Gusynin:2006gn,Gorbar:2007xh,Gorbar:2008hu,
Herbut:2008ui,Semenoff:2010zd,Semenoff:2011ya,Gorbar:2011kc}
and bilayer \cite{2010JETPL..91..314G,2010PhRvB..81o5451G,Gorbar:2011ce,Gorbar:2012jc}
graphene in the regime of the quantum Hall effect. Finally, the generalization of 
magnetic catalysis was also made to non-Abelian chromomagnetic fields  
\cite{Vshivtsev:1994si,Klimenko:1993ec,Shovkovy:1995td,Gusynin:1997vh,
Ebert:1997um,Ebert:2001bb,Zhukovsky:2001iq}, where the dynamics is 
dimensionally reduced by one unit of space, $D\to D-1$. 
For earlier reviews on magnetic catalysis, 
see Refs.~\cite{Miransky:1995rb,Gusynin:1999ti}.

\section{The Essence of Magnetic Catalysis}
\label{sec:2}

As already mentioned in the Introduction, the essence of magnetic catalysis is intimately 
connected with the dimensional reduction, $D\to D-2$, of charged Dirac fermions in the 
presence of a constant magnetic field. In this section, we discuss in detail how such a 
dimensional reduction appears and what implications it has for the spontaneous symmetry 
breaking. 

\subsection{Dimensional reduction in a magnetic field}

Before considering a fully interacting theory and all details of the dynamics responsible 
for the generation of the chiral condensate and the symmetry breaking, associated with it,
let us start from a free Dirac theory in a constant external magnetic field. This appears to 
be a perfect setup to understand the kinematic origin of the dimensional reduction, $D\to D-2$.

\subsubsection{Dirac fermions in a magnetic field in $3+1$ dimensions}
 
Let us start by reviewing the spectral problem for charged $(3+1)$-dimensional Dirac 
fermions in a constant magnetic field. We assume that the field is pointing in the positive 
$x^3$-direction. The corresponding Lagrangian density reads
\begin{equation}
{\cal L}= \bar{\Psi}(i\gamma^\mu D_\mu-m)\Psi ,
\label{DiracLagrangian}
\end{equation}
where the covariant derivative $D_{\mu} =\partial_{\mu}-i e A^{\rm ext}_{\mu}$
depends on the external gauge field. Without loss of generality, the external field 
$A^{\rm ext}_{\mu}$ is taken in the Landau gauge, 
$A_\mu^{\rm ext}\equiv (0,-\mathbf{A}^{\rm ext})$, where  
\begin{equation}
\mathbf{A}^{\rm ext}=\left(0,B x^1 ,0\right),
\label{extpotential}
\end{equation}
and $B$ is the magnetic field strength. By solving the Dirac equation of motion, one finds 
the following energy spectrum of fermions 
\cite{Akhiezer:1965}:
\begin{equation}
E_n^{(3+1)}(p_3) = \pm\sqrt{m^2+2|eB|n+(p_3)^2} ,
\label{spectrum3+1}
\end{equation}
where $n=0,1,2,\dots$ is the Landau level index. It should be noted that the Landau level 
index $n$ includes orbital and spin contributions: $n \equiv k+s+\frac{1}{2}$, where $k=0,1,2,\dots$ 
is an integer quantum number associated with the orbital motion, while $s=\pm \frac{1}{2}$ 
corresponds to the spin projection on the direction of the field. [For the orbital part of the wave 
functions, see Eq.~(\ref{wave-fun-orbital}) in the Appendix.] Considering that the energy depends 
only on $n$, we see that the energy of a quasiparticle in orbital state $k$ and spin $s=+\frac{1}{2}$
is degenerate with the energy of a quasiparticle in orbital state $k+1$ and spin $s=-\frac{1}{2}$. 
The lowest Landau level with $n=0$ is special: 
it corresponds to the lowest orbital state $k=0$ and has only one spin projection $s=-\frac{1}{2}$. 
The letter, in particular, implies that the lowest Landau level is a spin polarized state.

On top of the spin degeneracy of higher Landau levels ($n>0$), there is an additional (infinite)
degeneracy of each level with a fixed $n$ and a fixed value of the longitudinal momentum 
$p_3$. It is connected with the momentum $p_2\in \mathbb{R}$, which is a good quantum 
number in the Landau gauge utilized here. As follows from the form of the orbital wave 
functions in Eq.~(\ref{wave-fun-orbital}), the value of $-p_2/|eB|$ also determines the location 
of the center of a fermion orbit in the $x^1$-direction. A simple analysis \cite{Akhiezer:1965} 
shows that the area density of such states in the perpendicular $x^1 x^2$-plane is 
$\frac{|eB|}{2\pi}$ for $n=0$ and $\frac{|eB|}{\pi}$ for $n>0$ (here the double spin 
degeneracy of the higher Landau levels is accounted for). 

When the Dirac mass is much smaller than the corresponding magnetic energy scale 
(i.e., $m \ll\sqrt{|eB|}$), we find that the low-energy sector of the Dirac theory is determined
exclusively by the lowest Landau level ($n=0$). As we see from Eq.~(\ref{spectrum3+1}), 
the corresponding spectrum of the low-energy excitations is given by $E(p_3) = \pm\sqrt{m^2+(p_3)^2}$,
which is identical to the spectrum of a $(1+1)$-dimensional quantum field theory with a 
single spatial coordinate, identified with the longitudinal direction. This spectrum of the low-energy 
theory confirms the obvious kinematic aspect of the dimensional reduction, $3+1 \to 1+1$, 
in a constant magnetic field.

From the physics viewpoint, the dimensional reduction is the result of a partially restricted  
motion of Dirac particles in the $x^1 x^2$-plane perpendicular to the magnetic field. The  
effect can be seen already at the classical level in the so-called cyclotron motion, when the Lorentz 
force causes charged particles to move in circular orbits in $x^1 x^2$-plane, but does not 
constrain their motion along the $x^3$-direction. A very important new feature at the quantum 
level is the quantization of perpendicular orbits. Without such a quantization, the clean separation 
of the low-energy sector, dominated exclusively by the lowest Landau level, would not be 
possible.  

It should be noted that the spin also plays an important role in the dimensional reduction 
of Dirac particles. If the spin contribution were absent ($s=0$), the energy of the lowest Landau level 
would scale like $\sqrt{|eB|}$, which is not vanishingly small compared to 
the energy of the next Landau level $\sqrt{3|eB|}$. Then, a clean separation of the 
lowest Landau level into a dimensionally reduced, low-energy sector of the theory would become 
unjustified and meaningless.

\subsubsection{Dirac fermions in a magnetic field in $2+1$ dimensions}

It is straightforward to obtain the spectrum of charged Dirac fermions also in $2+1$ dimensions.
The vector potential in the Landau gauge takes the form: $\mathbf{A}^{\rm ext}=\left(0,B x^1\right)$. 
In the absence of the longitudinal direction $x^3$, the magnetic field $B$ is not an axial vector, 
but a pseudo-scalar. Concerning the Dirac algebra in $2+1$ dimensions, there exist two 
inequivalent irreducible representations, given by
\begin{equation}
\gamma^0=\sigma_3, 
\quad 
\gamma^1=i\sigma_1,
\quad 
\gamma^2=i\sigma_2 ,
\label{eq:pauli}
\end{equation}
and
\begin{equation}
\gamma^0=-\sigma_3, 
\quad 
\gamma^1=-i\sigma_1, 
\quad 
\gamma^2=-i\sigma_2 ,
\label{eq:pauli1}
\end{equation}
where $\sigma_i$ are the Pauli matrices. In each of these representations, the nature of the 
lowest Landau level is somewhat unusual:
it has either only a particle state (with a positive energy $E_0=m$) or only an antiparticle state
(with a negative energy $E_0=-m$). Such an asymmetry in the spectrum is known to induce a 
Chern-Simons term in the gauge sector of the theory \cite{Niemi:1983rq,Redlich:1983kn}. In order 
to avoid the unnecessary complication, it is convenient to use the following reducible 
representation instead:
\begin{equation}
\gamma^0= \left(\begin{array}{cc} \sigma_3 & 0 \\ 0&
-\sigma_3\end{array}\right) , \quad 
\gamma^1= \left(\begin{array}{cc}
i\sigma_1&0\\0&-i\sigma_1\end{array}\right), \quad 
\gamma^2= \left(\begin{array}{cc}i\sigma_2&0\\0&-i\sigma_2\end{array}\right).
\label{gamma2+1-reducible}
\end{equation}
(Incidentally, in the low-energy 
theory of  graphene, which is a real quasi-relativistic system in $2+1$ dimensions, such 
a reducible representation appears automatically \cite{Semenoff:1984dq}.) The corresponding 
Dirac spectrum reads
\begin{equation}
E_n^{(2+1)} = \pm\sqrt{m^2+2|eB|n} .
\label{spectrum2+1}
\end{equation}
As we see, this is very similar to the $(3+1)$-dimensional result in Eq.~(\ref{spectrum3+1}), 
except for the missing dependence on the longitudinal momentum $p_3$.

Repeating the same arguments as in the $(3+1)$-dimensional case, we find that the low-energy 
sector of the Dirac theory in $2+1$ dimensions is also determined by the lowest Landau level. 
Here we assume again that the Dirac mass is much smaller than the corresponding Landau energy 
scale ($m \ll \sqrt{|eB|}$) in order to insure a clear separation of the low- and 
high-energy scales. 

Just like in the higher dimensional case, all Landau levels are (infinitely) degenerate. In particular, 
the number of degenerate states per unit area is $\frac{|eB|}{2\pi}$ in the lowest Landau level. 
A special feature of the $(2+1)$-dimensional theory is a discrete, rather than continuous 
spectrum of excitations. In the absence of the $x^3$-direction and the associated quantum 
number $p_3$, all positive energy states in the lowest Landau level have the same energy 
$E_0=m$. Moreover, when $m\to0$, this energy goes to zero and becomes degenerate with 
the negative energy states $E_0=-m$. In this limit, there is an infinite vacuum degeneracy 
even if the condition of charge neutrality may favor a state with exactly half-filling of the 
lowest Landau level. It should be expected, however, that taking into account any type of 
fermion interaction will lead to a well defined ground state, in which the interaction energy 
is minimized. One can even make an educated guess that the corresponding ground state 
should be a Mott-type insulator with a dynamically generated mass/gap. 

Another special and rather unusual feature of the $(2+1)$-dimensional Dirac fermions in a 
magnetic field is a spontaneous symmetry breaking, which is manifested by a nonzero ``chiral" 
condensate $\langle \bar\Psi \Psi \rangle$ already in the {\em free} theory. To see this, let 
us make use of the proper-time representation of the fermion propagator in the magnetic 
field, see Eq.~(\ref{proper-time-2+1}) in the Appendix. In the limit of a small bare mass ($m_0\to 0$), we
easily derive the following (regularized) expansion for the condensate: 
\begin{eqnarray}
\langle \bar\Psi \Psi \rangle &\equiv& -  \mbox{tr}\left[S_{2+1}(x,x)\right]
= - \frac{m_0 |eB|}{2\pi^{3/2}}\int_{1/\Lambda^2}^{\infty}\frac{ds}{\sqrt{s}}e^{-s m_0^2}\coth(s|eB|)
\nonumber\\
&\simeq& -\frac{|eB|}{2\pi} \mbox{sign}(m_0)
-\frac{m_0}{\pi^{3/2}} \left[ \Lambda  +\sqrt{\frac{\pi |eB|}{2}} \zeta\left(\frac{1}{2},1+\frac{m_0^2}{2|eB|}\right)\right].
\label{condensate2+1}
\end{eqnarray}
It can be shown that the first term, which remains nonzero even in the massless limit, comes from 
the lowest Landau level. At first sight, this may appear to be a very surprising result. Upon a closer examination, 
one finds that this condensate is directly connected with a nonzero density of states and a nonzero
spin polarization in the lowest Landau level of the free Dirac theory. The result is 
unambiguous only after specifying the sign of the bare mass parameter, which is also typical 
for spontaneous symmetry breaking. 

In connection with the result in Eq.~(\ref{condensate2+1}), it may be useful to recall that the 
chirality is not well defined in the $(2+1)$-dimensional space. However, as we will discuss in 
Sec.~\ref{mag-cat-2D}, the condensate $\langle \bar\Psi \Psi \rangle$ is still of interest 
because it breaks another global symmetry that has a status similar to that of the conventional 
chiral symmetry.

\subsection{Magnetic catalysis in 2+1 dimensions}
\label{mag-cat-2D}

Now, let us consider a Nambu-Jona-Lasinio (NJL) model in $2+1$ dimensions, 
in which the magnetic catalysis of symmetry breaking is realized in its simplest 
possible form \cite{Gusynin:1994re,Gusynin:1994va}. When using the reducible
representation of Dirac algebra, given by Eq.~(\ref{gamma2+1-reducible}), one 
finds that the kinetic part of the massless Dirac theory is invariant under a global
$U(2)$ flavor symmetry. The generators of the symmetry transformations are 
given by $T_0=I$, $T_1=\gamma^5,T_2=\frac{1}{i}\gamma^3$, and 
$T_3=\gamma^3\gamma^5$, where 
$\gamma^5\equiv i\gamma^0\gamma^1\gamma^2\gamma^3$.
A dynamical Dirac mass will break this $U(2)$ symmetry down to the
$U(1)\times U(1)$ subgroup with generators $T_0$ and $T_3$. 

The NJL-type Lagrangian density, with the interaction term invariant 
under the $U(2)$ flavor symmetry, can be written down as follows:
\begin{equation}
{\cal L} = \bar{\Psi}\, i\gamma^\mu D_\mu \Psi +
\frac{G}{2} \left[ (\bar{\Psi}\Psi)^2+(\bar{\Psi}i\gamma^5\Psi)^2 +
(\bar{\Psi}\gamma^3\Psi)^2\right],
\label{LagrangianNJL}
\end{equation}
where $G$ is a dimensionfull coupling constant. This theory is nonrenormalizable,
but can be viewed as a low-energy effective theory with a range of validity 
extending up to a certain ultraviolet energy scale set by a physically motivated choice 
of the cutoff parameter $\Lambda$. 
  
\subsubsection{Weak coupling approximation}

As usual in problems with spontaneous symmetry breaking, we use the method 
of Schwinger-Dyson (gap) equation in order to solve for the dynamical mass 
parameter. We assume that the structure of the (inverse) full fermion propagator 
is the same as in the free theory, but has a nonzero dynamical Dirac mass $m$,
\begin{equation}
S^{-1}(x,x^\prime)
= -i \left[ i \gamma^0 \partial_t - (\bm{\pi} \cdot\bm{\gamma})-m\right]
\delta^{3}(x-x^\prime).
\label{prop-S1}
\end{equation}
In the mean-field approximation, the dynamical mass parameter satisfies 
the following gap equation:
\begin{equation}
m= G\,  \mbox{tr}\left[S(x,x)\right] .
\label{m-small-G}
\end{equation}
To leading order in weak coupling, $G\to 0$, this equation can be solved perturbatively.
Indeed, by substituting the condensate calculated at $m_0\to 0$ in the {\em free} theory, see 
Eq.~(\ref{condensate2+1}), into the right-hand side of the gap equation (\ref{m-small-G}), 
we obtain
\begin{equation}
|m| \simeq G\,  \frac{|eB|}{2\pi}.
\label{perturb-solution-2+1}
\end{equation}
As we see, the dynamical mass is induced at any nonzero attractive 
coupling. 

The massless NJL model in Eq.~(\ref{LagrangianNJL}) is invariant under the flavor $U(2)$ symmetry. 
The generation of a Dirac mass $m$ is only one of many equivalent ways of breaking this 
symmetry down to a $U(1)\times U(1)$ subgroup. Indeed, by applying a general $U(2)$ transformation, 
we find that the Dirac mass term can be turned into a linear combination of the following 
three mass terms: $m$, $\gamma^3 m_3$, and $i\gamma^5 m_5$. (In principle, 
there is also a possibility of the so-called Haldane mass term $\gamma^3 \gamma^5 \Delta$, 
which is a singlet under $U(2)$. We do not discuss it here. However, as we will see 
in Sec.~\ref{sec:3.3}, the Haldane mass plays an important role in graphene.)

In our perturbative analysis, we did not get any nonzero $m_3$ or $m_5$ because the vacuum alignment 
was predetermined by a ``seed" Dirac mass $m_0$ in the free theory, see Eq.~(\ref{condensate2+1}).

\subsubsection{Large $N$ approximation}

It is instructive to generalize the above analysis in the NJL model to the case of strong coupling. 
While magnetic catalysis occurs even at arbitrarily weak coupling, such a generalization will 
be useful to understand how magnetic catalysis is lost in the limit of the vanishing magnetic field. 

At strong coupling, a reliable solution to the NJL model can be obtained by using the so-called 
large $N$ approximation, which is rigorously justified when the fermion fields in Eq.~(\ref{LagrangianNJL}) 
carry an additional, ``color" index $\alpha=1, 2, \ldots,N$, and $N$ is large. Using the 
Hubbard-Stratonovich transformation \cite{1957SPhD....2..416S,1959PhRvL...3...77H}, 
one can show that the NJL theory in Eq.~(\ref{LagrangianNJL}) is equivalent to the  following one:
\begin{equation}
{\cal L}=  \bar{\Psi}\, i\gamma^\mu D_\mu  \Psi 
 - \bar{\Psi}(\sigma+\gamma^3\tau+i\gamma^5\pi)\Psi 
 - \frac{1}{2G} \left(\sigma^2+\pi^2+\tau^2\right).
\label{eq:11}
\end{equation}
Note that the equations of motion for the new composite fields read
\begin{equation}
\sigma = - G(\bar{\Psi}\Psi), \quad
\tau = -G(\bar{\Psi}\gamma^3\Psi),\quad
\pi = -G (\bar{\Psi}i\gamma^5\Psi).
\end{equation}  
Under $U(2)$ flavor symmetry transformations, these composite fields transform into linear 
combinations of one another, but the quantity $\sigma^2+\pi^2+\tau^2$ remains invariant. 

The effective action for the composite fields,
\begin{equation}
\Gamma = -\frac{1}{2G}\int d^3x(\sigma^2+\tau^2+\pi^2) 
- i \, \mbox{tr}\,\mbox{Ln} \left[i\gamma^\mu D_\mu
- (\sigma+\gamma^3\tau+i\gamma^5\pi)\right],    \label{eq:13}
\end{equation}
is obtained by integrating out the fermionic degrees of freedom from the action.
It is convenient to expand this effective action in powers of derivatives of the composite 
fields. The leading order in such an expansion is the effective potential $V$ (up to the
minus sign). Because of the flavor symmetry, the effective potential depends
on $\sigma$, $\pi$, and $\tau$ fields only through their $U(2)$ invariant combination 
$\rho^2=\sigma^2+\pi^2+\tau^2$.

Using the proper-time regularization, one obtains the following explicit expression
for the effective potential \cite{Gusynin:1994re,Gusynin:1994va}:
\begin{equation}
V(\rho)\simeq \frac{N}{\pi} \Bigg[\frac{\Lambda}{2}
\left(\frac{1}{g}-\frac{1}{\sqrt{\pi}}\right) \rho^2- \sqrt{2}|eB|^{3/2}\zeta
\left(-\frac{1}{2}, 1+ \frac{\rho^2}{2|eB|} \right) 
-  \frac{\rho|eB|}{2}\Bigg] ,     
\label{eq:poten}
\end{equation}
where we dropped the terms suppressed by the ultraviolet cutoff parameter $\Lambda$
and introduced a dimensionless coupling constant $g\equiv N \Lambda G/\pi $.

The field configuration $\rho$ that minimizes the effective potential is determined by 
solving the equation $dV/d\rho=0$, i.e., 
\begin{equation}
\Lambda \left(\frac{1}{g}-\frac{1}{\sqrt{\pi}}\right)\rho = \frac{|eB|}{2}+\rho\sqrt{\frac{|eB|}{2}}
\zeta\left(\frac{1}{2},1+\frac{\rho^2}{2|eB|}\right).
\label{gap-eq-17}
\end{equation}
In essence, this is the gap equation.
At weak coupling, $g\to0$, in particular, we obtain the following approximate solution:
\begin{equation}
m=\rho_{\rm min} \simeq G N\frac{|eB|}{2\pi },
\label{m-small-G-large-N}
\end{equation}
which is the large $N$ generalization of the result for the dynamical mass in Eq.~(\ref{perturb-solution-2+1}).

\subsubsection{Zero magnetic field limit in $2+1$ dimensions}

Before concluding the discussion of the $(2+1)$-dimensional NJL model, it is instructive 
to consider how the above analysis of flavor symmetry breaking modifies in the zero 
magnetic field case. By taking the limit $B\to 0$ in Eq.~(\ref{eq:poten}), 
we arrive at the following effective potential:
\begin{equation}
V_{B=0}(\rho)\simeq \frac{N}{\pi} \Bigg[\frac{\Lambda}{2}
\left(\frac{1}{g}-\frac{1}{\sqrt{\pi}}\right)\rho^2
+\frac{1}{3}\rho^3\Bigg].
\label{eq:potenB0}
\end{equation}
Now, the zero-field limit of the gap equation, $dV_{B=0}/d\rho=0$, is given by
\begin{equation}
\Lambda \left(\frac{1}{\sqrt{\pi}}-\frac{1}{g}\right)\rho =\rho^2.
\end{equation}
This is very different from Eq.~(\ref{gap-eq-17}). In particular, the only solution to this 
equation at $g< \sqrt{\pi}$ is a trivial one, $\rho=0$. The nontrivial solution appears 
only in the case of a sufficiently strong coupling constant, $g>\sqrt{\pi}$. This is in a 
stark contrast with the dynamical mass generation in the presence of a magnetic field, 
where a nontrivial solution exists at arbitrarily small values of the coupling constant $g$.

\subsection{Magnetic catalysis in 3+1 dimensions}
\label{mag-cat-3D}

Let us now extend the analysis of the previous subsection to the case of a 
$(3+1)$-dimensional model, where the dynamics is truly nonperturbative. 
The Lagrangian density of the corresponding NJL model reads
\begin{equation}
{\cal L} =  \bar{\Psi} \, i\gamma^\mu D_\mu \Psi +
\frac{G}{2} \left[ (\bar{\Psi}\Psi)^2+(\bar{\Psi}i\gamma^5\Psi)^2 \right].
\label{LagrangianNJL3+1}
\end{equation}
This model possesses the $U(1)_L\times U(1)_R$ chiral symmetry. The symmetry 
will be spontaneously broken down to the $U(1)_{L+R}$ subgroup 
when a dynamical Dirac mass is generated.

\subsubsection{Weak coupling approximation}

In the weakly coupled limit, the gap equation in the mean-field approximation  reads
\begin{equation}
m= G\,  \mbox{tr}\left[S(x,x)\right] .
\label{m-small-G-3+1}
\end{equation}
Formally, it is same as the gap equation in the $(2+1)$-dimensional model in 
Eq.~(\ref{m-small-G}). However, as we will see now, its symmetry breaking 
solution will be qualitatively different. 

Let us start by showing that the chiral condensate vanishes in the free theory in $3+1$ 
dimensions when the bare mass goes to zero, $m_0\to 0$. By making use of the proper-time 
representation, see Eq.~(\ref{proper-time3+1}) in the Appendix, we obtain
\begin{eqnarray}
\langle \bar\Psi \Psi \rangle &\equiv& -  \mbox{tr}\left[S(x,x)\right]
=-\frac{m_0 |eB|}{(2\pi)^2}\int_{1/\Lambda^2}^{\infty}\frac{ds}{s}e^{-s m_0^2}\coth(s|eB|)
\nonumber\\
\simeq && \hspace{-10pt}
-\frac{m_0}{(2\pi)^2}\left[\Lambda^2-m_0^2\left(\ln\frac{\Lambda^2}{2|eB|}-\gamma_E\right)
+|eB|\ln\frac{m_0^2}{4\pi |eB|}+2|eB|\ln\Gamma\left(\frac{m_0^2}{2 |eB|}\right)\right]\nonumber\\
\simeq &&\hspace{-10pt}
-\frac{m_0}{(2\pi)^2}\left[\Lambda^2
+|eB|\ln\frac{|eB|}{\pi m_0^2} - m_0^2 \ln\frac{\Lambda^2}{2|eB|} +O\left(\frac{m_0^4}{|eB|}\right)\right].
\label{condensate3+1}
\end{eqnarray}
In the limit $m_0\to 0$, we see that the condensate indeed vanishes. This means that 
we cannot apply a perturbative approach to find any nontrivial (symmetry breaking) 
solutions to the gap equation in Eq.~(\ref{m-small-G-3+1}).

The explicit form of the gap equation (\ref{m-small-G-3+1}) reads
\begin{equation}
m \simeq G\, \frac{m}{(2\pi)^2}\left[\Lambda^2 +|eB|\ln\frac{|eB|}{\pi m^2}  \right] .
\label{gap-eq-3+1}
\end{equation}
Its nontrivial solution is given by 
\begin{equation}
m\simeq \sqrt{\frac{|eB|}{\pi}}\exp\left(\frac{\Lambda^2}{2|eB|}\right)
\exp\left(-\frac{2\pi^2}{G|eB|}\right).
\label{dyn-mass-nonpertur}
\end{equation}
When $G\to 0$ this result reveals an essential singularity. Obviously, such a dependence 
cannot possibly be obtained by resuming any finite number of perturbative corrections in 
powers of a small coupling constant. Therefore, despite the weak coupling, the result
for the dynamical mass (\ref{dyn-mass-nonpertur}) is truly nonperturbative.

\subsubsection{Zero magnetic field limit  in $3+1$ dimensions}

It is instructive to compare the above dynamics of spontaneous symmetry breaking with 
case of the zero magnetic field. At $B=0$, the chiral condensate in the free theory is easily 
obtained by taking the appropriate limit in Eq.~(\ref{condensate3+1}), i.e.,
\begin{equation}
\langle \bar\Psi \Psi \rangle_{B=0} = -\frac{m_0}{(2\pi)^2}\left[\Lambda^2
-m_0^2\left(\ln\frac{\Lambda^2}{m_0^2}+1-\gamma_E\right)\right].
\end{equation}
The corresponding gap equation is 
\begin{equation}
m \simeq G\, \frac{m}{(2\pi)^2}\left[\Lambda^2 -m^2 \ln\frac{\Lambda^2}{m^2}  \right] .
\label{gap-eq-3+1_B=0}
\end{equation}
Because of the negative sign in front of the logarithmic term, this equation does not
have any nontrivial solutions for the dynamical mass at vanishingly small coupling 
constant $g\equiv G\Lambda^2/(2\pi)^2$. In fact, for the whole range of subcritical 
values, $g<1$, the only solution to this gap equation is $m=0$. The nontrivial solution 
appears only in the case of sufficiently strong coupling, $g>1$.

\subsection{Symmetry breaking as bound state problem}
\label{sec:2.4}

In this subsection, we consider an alternative approach to the problem of chiral
symmetry breaking in the NJL model in a constant magnetic field. As we will see, 
this approach is particularly beneficial for illuminating the role of the dimensional 
reduction in magnetic catalysis. 

Instead of solving the gap equation, we consider the problem of bound states with 
the quantum numbers of the Nambu-Goldstone bosons, using the method of a 
homogeneous Bethe-Salpeter equation (for a review, see 
Ref.~\cite{Miransky:1994vk}). The underlying idea for this framework is 
motivated by the Goldstone theorem \cite{Nambu:1960xd,Goldstone:1961eq,Goldstone:1962es}.
The theorem states that spontaneous breaking of a continuous 
global symmetry leads to the appearance of new massless scalar particles 
(i.e., Nambu-Goldstone bosons) in the low-energy spectrum of the theory. The total 
number of Nambu-Goldstone bosons and their quantum numbers are determined by the 
broken symmetry generators. 

The homogeneous Bethe-Salpeter equation for a pion-like state
takes the form \cite{Miransky:1994vk,Gusynin:1995nb}:
\begin{eqnarray}
\chi_{ab}(x,y;P) &=& -i\int d^4x^{\prime}d^4y^{\prime}d^4x^{\prime\prime}
d^4y^{\prime\prime}S_{aa_1}(x,x^{\prime})
K_{a_1b_1;a_2b_2}(x^{\prime}y^{\prime},x^{\prime\prime}y^{\prime\prime}) \nonumber\\
&\times& \chi_{a_2b_2}(x^{\prime\prime},y^{\prime\prime};P)
S_{b_1b}(y^{\prime},y),
\label{BS-equation}
\end{eqnarray}
where $\chi_{ab}(x,y;P)\equiv \langle 0|T\psi_a(x)\bar\psi_b(y)|P;\pi\rangle$ 
is the Bethe-Salpeter wave function of the bound state boson with 
four-momentum $P$, 
and $S_{ab}(x,y)=\langle 0|T\psi_a(x)\bar\psi_b(y)|0\rangle$ is the fermion
propagator. Here and below, the sum over repeated Dirac indices ($a_1$, 
$b_1$, $a_2$, $b_2$) is assumed. The explicit form the Bethe-Salpeter 
kernel is \cite{Miransky:1994vk,Gusynin:1995nb}:
\begin{eqnarray}
K_{a_1b_1;a_2b_2}(x^{\prime}y^{\prime},x^{\prime\prime}y^{\prime\prime}) &=& 
G\big[
   \delta_{a_1b_1}\delta_{b_2a_2}
+ (i\gamma_5)_{a_1b_1} (i\gamma_5)_{b_2a_2} 
- (i\gamma_5)_{a_1a_2}  (i\gamma_5)_{b_2b_1}  \nonumber\\
&&-\delta_{a_1a_2}\delta_{b_2b_1} \big]  
\delta^4(x^{\prime}-y^{\prime})
\delta^4(x^{\prime}-x^{\prime\prime})
\delta^4(x^{\prime}-y^{\prime\prime}) . 
\label{kernel}
\end{eqnarray}
It is convenient to rewrite the wave function in terms of the relative coordinate 
$z\equiv x-y$ and the center of mass coordinate $X\equiv (x+y)/2$,
\begin{equation}
\chi_{ab}(X,z;P) =  e^{i s_\perp X^1 z^2/\ell^2 } e^{-iP_\mu X^\mu} \tilde\chi_{ab}(X,z;P) ,
\label{chi-ansatz}
\end{equation}
where we factorized the Schwinger phase factor, see Eq.~(\ref{schwinger-phase}),
and introduced the notation:  $s_{\perp} \equiv \mbox{sign}(eB)$ and 
$\ell=1/\sqrt{|e B |}$. After substituting 
the wave function (\ref{chi-ansatz}) and the kernel (\ref{kernel}) into
Eq.~(\ref{BS-equation}), we arrive at the following equation:
\begin{eqnarray}
\tilde\chi_{ab}(z;P)&=&-i G \int d^4 X^\prime
\tilde S_{aa_1}\left(\frac{z}{2}-X^{\prime}\right)
\Big[\delta_{a_1b_1}~\mbox{tr}\left[\tilde\chi(0;P)\right] 
-(\gamma_5)_{a_1b_1}\mbox{tr}\left[\gamma_5\tilde\chi(0;P)\right]\nonumber\\
&&
-\tilde\chi_{a_1b_1}(0;P) +
(\gamma_5)_{a_1a_2}\tilde\chi_{a_2b_2}(0;P) 
(\gamma_5)_{b_2b_1}\Big] \tilde{S}_{b_1b}\left(\frac{z}{2}+X^{\prime}\right)
\nonumber\\
&& \times 
e^{i\frac{s_\perp}{2\ell^2}\left(z^1 X^{\prime 2}-X^{\prime 1}z^2\right)}
e^{iP_\mu X^{\prime\mu}} .
\label{Bethe-Salpeter-coordinate}
\end{eqnarray}
Here we took into account that the equation admits a translation invariant solution 
and replaced $\tilde\chi_{ab}(X,z;P)\to \tilde\chi_{ab}(z,P)$. Note that, on the right-hand side
of Eq.~(\ref{Bethe-Salpeter-coordinate}), the dependence on the center of mass coordinate $X$ completely 
disappeared after a shift of the integration variable $X^{\prime}\to X^{\prime}-X$ was 
made. 

In the lowest Landau level approximation, one can show that the Fourier transform of 
the Bethe-Salpeter wave function takes the form \cite{Gusynin:1995nb}: 
\begin{equation}
\tilde\chi_{ab}(p;P\to 0) = A(p_\parallel) e^{-p_\perp^2\ell^2} 
\frac{\gamma^0\omega-\gamma^3p^3-m}{\omega^2-(p^3)^2-m^2}
\gamma^5 {\cal P}_{+}
\frac{\gamma^0\omega-\gamma^3p^3-m}{\omega^2-(p^3)^2-m^2},
\end{equation}
where $p_\parallel =(\omega,p^3)$, $p_\perp^2=(p^1)^2+(p^2)^2$, and 
${\cal P}_{\pm} = \frac{1}{2}\left(1\pm i s_{\perp} \gamma^1\gamma^2\right)$. 
The new function $A(p_\parallel)$ satisfies 
the equation:
\begin{equation}
A(p_\parallel)=\frac{G|eB|}{4\pi^3} \int  \frac{A(k_\parallel) d^2 k_\parallel}{k_\parallel^2+m^2},
\label{BS-A}
\end{equation}
where we made the Wick rotation ($\omega \to i \omega$). The solution to this equation is a
constant: $A(p_\parallel)=C$. Dropping nonzero $C$ and cutting off the integration at $\Lambda$, 
we finally arrive at the gap equation for the mass parameter $m$:
\begin{equation}
1 \simeq \frac{G|eB|}{4\pi^2} \int^{\Lambda^2}_0 \frac{dk_\parallel^2}{k_\parallel^2+m^2}.
\label{gap-eq-37}
\end{equation}                         
The solution to this equation is
\begin{equation}
m \simeq \Lambda^2\exp\left(-\frac{2\pi^2}{G|eB|}\right) ,
\label{dyn-mass-nonpertur-35}
\end{equation}
which, to leading order, agrees with the solution obtained  in 
Eq.~(\ref{dyn-mass-nonpertur}). 

The Bethe-Salpeter equation (\ref{BS-A}) can be rewritten in the form of a two-dimen\-sional 
Schr\"odinger equation with an attractive $\delta$-function potential. In order to see this explicitly, 
let us introduce the following wave function
\begin{equation}
\psi(\mathbf{r})=\int \frac{d^2k_\parallel}{(2\pi)^2}
\frac{e^{-i{\bf k_\parallel r}}}{k_\parallel^2+m^2} A(k_\parallel).
\label{wave-function}
\end{equation}
Taking into account that $A(p_\parallel)$ satisfies Eq.~(\ref{BS-A}), it is straightforward to 
show that the wave function $\psi(\mathbf{r})$ satisfies the following Schr\"odinger type equation:
\begin{equation}
\left(-\frac{\partial^2}{\partial r^2_1}-
\frac{\partial^2}{\partial r^2_2} +m^2-\frac{G|eB|}{\pi}
\delta^2_\Lambda(\mathbf{r})\right)\psi(\mathbf{r})=0,
\label{eq-Schrodinger-37}
\end{equation}
in which $-m^2$ plays the role of the energy $E$. Since $m^2$ must be positive, 
the problem is reduced to finding the spectrum of bound states (with $E=-m^2<0$) in the 
Schr\"odinger problem. The potential energy in Eq.~(\ref{eq-Schrodinger-37}) is 
expressed in terms of
\begin{equation}
\delta^2_\Lambda(\mathbf{r})=\int_\Lambda
\frac{d^2k_\parallel}{(2\pi)^2} e^{-i{\bf k_\parallel r}} ,
\end{equation}
which is a regularized version of the $\delta$-function that describes the local interaction in the 
NJL model. 

Notice that, by using the same approach, one can show that the Bethe-Salpeter equation for a massless
NG-boson state in the NJL model in $2+1$ dimensions can be reduced to the gap equation
\begin{equation}
A(p)=\frac{G|eB|}{2\pi^2} \int^\Lambda_{-\Lambda} \frac{A(k)dk}{k^2+m^2}.
\end{equation}
This has the same solution for the mass as in Eq.~(\ref{perturb-solution-2+1}). Also, this integral 
equation is equivalent to the following one-dimensional Schr\"odinger equation:
\begin{equation}
\left(-\frac{d^2}{d x^2}+m^2 -
\frac{G|eB|}{\pi^2} \delta_\Lambda(x)\right)\psi(x)=0 ,
\end{equation}
where the regularized version of the $\delta$-function is given by 
$\delta_\Lambda(x)=\int^\Lambda_{-\Lambda} \frac{dk}{2\pi} e^{-ikx}$.

\subsection{Analogy with superconductivity}

It is interesting to point that the dynamics described by the gap equation in the case of 
magnetic catalysis has a lot of conceptual similarities to the mechanism of superconductivity in 
metals and alloys. This is despite the clear differences between the two phenomena that 
we discussed in the Introduction. (In order to avoid a possible confusion, let us emphasize 
that here we compare the nonrelativistic Cooper pairing dynamics in superconductivity in 
the {\em absence} of magnetic fields with the relativistic dynamical generation of a mass in 
the {\em presence} of a constant magnetic field.)

The corresponding gap equation in the Bardeen-Cooper-Schrieffer theory \cite{Bardeen:1957mv}
of superconductivity can be written in the following form:
\begin{equation}
1=G\,  N(0)  \int_{0}^{\hbar \omega_D}\frac{d\epsilon}{\sqrt{\epsilon^2+\Delta^2}},
\label{BCS}
\end{equation}
where $N(0)$ is the density of electron states at the Fermi surface, $\omega_D$ is the Debye 
frequency, and $\Delta$ is the energy gap associated with superconductivity. The solution for
the gap reads
\begin{equation}
\Delta \simeq \hbar \omega_D \exp\left(-\frac{1}{G\, N(0)}\right) .
\label{BCS-solution}
\end{equation}
At weak coupling, this solution has the same essential singularity as the dynamical mass 
parameter in Eq.~(\ref{dyn-mass-nonpertur-35}). We can argue that the similarity is not accidental. 
To see this clearly, let us rewrite the gap equation in the problem of magnetic catalysis in the lowest Landau level 
approximation, see Eq.~(\ref{gap-eq-37}), as follows:
\begin{equation}
1=G \frac{|eB|}{2\pi}\int \frac{d\omega d p^3}{\omega^2+(p^3)^2+m^2} 
\simeq G \frac{|eB|}{2\pi} \int_{0}^{\Lambda} \frac{d \omega} {\sqrt{\omega^2+m^2}},
\label{mag-cat}
\end{equation}
where the Wick rotation was performed ($\omega\to i\omega$). As we see, the structure of 
this gap equation is identical to its counterpart in the BCS theory after we identify the density 
of states $N(0)$ with the density of states in the lowest Landau level, $|eB|/(2\pi)$, and the Debye frequency 
$\omega_D$ with the cutoff parameter $\Lambda$. 

The similarity between the BCS theory of superconductivity and magnetic catalysis
goes deeper. In particular, the generation of a nonzero gap in superconductors can be 
also thought of as the result of a $3+1\to 1+1$ dimensional reduction of the phase space 
around the Fermi surface. Also, just like in magnetic catalysis, 
it is essential that the density of states at the Fermi surface is nonzero.

\subsection{Bound states in lower dimensions}

As we saw in Sec.~\ref{sec:2.4}, the problem of spontaneous symmetry breaking and 
the associated dynamical generation of the Dirac mass can be reformulated as a problem 
of composite massless states with the quantum numbers of Nambu-Goldstone bosons. 

In the presence of a constant magnetic field, in particular, we also found that the corresponding 
Bethe-Salpeter equation for the bound states can be recast in an equivalent form as a Schr\"odinger 
equation in a dimensionally reduced space. The dimensional reduction is $D\to D-2$ and, 
therefore, the relevant problem of bound states is considered in spaces of lower dimensions. 

In order to prove that the essence of magnetic catalysis is directly connected with this 
reduction, let us consider a simple quantum mechanical problem: the formation of bound 
states in a shallow potential well in spaces of various dimensions. As we will see, at least 
one bound state does exist in one- and two-dimensional cases \cite{Simon:1976cx,
Simon:1976un,1977AnPhy.108...69B,Landau:101811}, but not always in three dimensions. 
We will also see that, while the result for the binding energy is perturbative in the coupling 
constant in one dimension, it has an essential singularity in two dimensions.

\subsubsection{Bound states in a one-dimensional potential well}

Let us start from the simplest one-dimensional problem of a nonrelativistic particle of mass $m_{*}$ 
confined to move on a line. Let the potential energy of the well be given by $U(x)$, which is negative 
and quickly approaches zero when $|x|\to\infty$. One can show that even a vanishingly 
small depth of the potential well is sufficient to produce a bound state (i.e., a quantum state 
with a negative energy). The corresponding binding energy is given by \cite{Landau:101811}
\begin{equation}
|E_{\rm 1D}| \simeq \frac{m_{*}}{2\hbar^2}\left(-\int_{-\infty}^\infty U(x) dx \right)^2.
\end{equation}
If we rescale the potential energy $U(x)$ by a ``coupling 
constant" factor $g$, i.e., $U(x)\to g U(x)$, we find that $|E_{\rm 1D}| \sim g^2$ as $g\to 0$. 
In other words, the binding energy has a power-law dependence as a function of the depth 
of the potential energy $U(x)$. This is a typical result that can be obtained by perturbative 
techniques, controlled by powers of the small parameter $g$ \cite{1977AnPhy.108...69B}. 

The above conclusion remains valid basically for any attractive potential $U(x)$. For example, 
one can rigorously prove that, if $\int (1+|x|) |U(x)| dx< \infty$, there is a bound state for all 
small positive $g$ if and only if $\int U(x) dx \leq 0$ (i.e., the potential is attractive at least on 
average) \cite{Klaus:1977dz}.

\subsubsection{Bound states in a two-dimensional potential well}

In the case of a two-dimensional system (i.e., a nonrelativistic particle of mass $m_{*}$ 
confined to move on a plane), the general conclusion about the existence of a bound 
state around a potential well of a vanishingly small depth still remains valid. However, 
an important qualitative difference appears in the result. The binding energy reveals 
an essential singularity as a function of the depth of the potential well. In order to 
understand this better, let us consider a problem with a cylindrically symmetric potential 
energy $U(r)$, where $r$ is the radial polar coordinate in the plane. If the potential 
energy is sufficiently shallow and localized 
(i.e., $\left|\int_0^\infty r U(r) dr\right|\ll m_{*}/\hbar^2$), one finds that the energy 
of the bound state is given by \cite{Landau:101811}
\begin{equation}
|E_{\rm 2D}| \simeq \frac{\hbar^2}{m_{*} a^2}
\exp\left(-\frac{\hbar^2}{m_{*}}\left|\int_0^\infty r U(r) dr\right|^{-1}\right),
\end{equation}
where $a$ is the characteristic size of the potential well. The fact that this energy is 
singular can be made explicit by rescaling the potential energy $U(r)$: $U(r)\to g U(r)$. 
Then, we find that $|E_{\rm 2D}| \sim \exp(-C/g)$ as $g\to 0$ (here $C$ is a constant 
determined by the shape of the potential well). Unlike the $g^2$ power-law suppression 
of the binding energy in one dimension, this is a much stronger suppression indicating a much weaker
binding. Moreover and perhaps more importantly, such an essential singularity cannot possibly be 
obtained by resuming any finite number of perturbative corrections, controlled by powers 
of the small parameter $g$. Therefore, the singular behavior of the binding energy in two 
dimensions is a sign of a truly nonperturbative (albeit weakly-interacting) physics. 

Again, this result is very general. It can be rigorously proven that, in the case when $\int |U(x)|^{1+\epsilon} d^2x< \infty$ 
(with some $\epsilon>0$) and $\int (1+x^2)^{\epsilon} |U(x)| d^2 x< \infty$, there is a bound state for all 
small positive $g$ if and only if $\int U(x) d^2x \leq 0$ (i.e., the potential is attractive at least on 
average) \cite{Simon:1976cx,Simon:1976un}.

\subsubsection{Bound states in a three-dimensional potential well}

Now, in the three-dimensional case, there are no bound states if the potential well is too 
shallow in depth. This was first shown by Peierls in 1929 \cite{Peierls:1929fk}. This can 
be demonstrated, for example, in a special case of a spherically symmetric potential well 
of a finite size, 
\begin{equation}
U(r) = \left\{
\begin{array}{lll}
-g \frac{\pi^2 \hbar^2}{8m_{*} a^2} &\quad \mbox{for} \quad  & r\leq a,\\
0 &\quad  \mbox{for} \quad  & r>a.\\
\end{array}
\right. 
\end{equation}
The condition to have at least one bound state is $g>1$ \cite{Landau:101811}. In other words, the depths 
of the potential well (or the strength of the ``coupling constant" g) should be larger than the critical value, 
given by $g_{\rm cr}=1$. In the supercritical regime, $g=1+\epsilon$ with $0<\epsilon\ll 1$, the binding 
energy is given by \cite{Landau:101811}
\begin{equation}
|E_{\rm 3D}| = \frac{\pi^4 \hbar^2}{2^7 m_{*} a^2}\epsilon^2.
\end{equation}
In the subcritical regime $g<1$, on the other hand, there are no bound states at all.

\section{Magnetic Catalysis in Gauge Theories}

Motivated by the fact that magnetic catalysis has a rather general underlying 
physics, explained by the dimensional reduction of the particle-antiparticle 
pairing, it is natural to ask how it is realized in gauge theories with long-range 
interactions, such as QED. This problem was discussed in numerous studies 
\cite{Gusynin:1995gt,Leung:1995mh,
Parwani:1994an,Parwani:1995am,Gusynin:1997kj,Gusynin:1998zq,Farakos:1998ds,Lee:1998sr,
Gusynin:1999pq,Farakos:1999qc,Gusynin:1998nh,Gusynin:2000tv,Alexandre:2000yf,Alexandre:2001vu,
Gusynin:2003dz,Sadooghi:2007ys,Ayala:2009fv,Ayala:2010fm}. 
Here we will briefly review only the key 
results and refer the reader to the original papers for further details.

\subsection{Magnetic Catalysis in QED}
\label{sec:3.1}

Using the same conceptual approach as outlined in Sec.~\ref{sec:2.4} for the 
NJL model, one can show that, in Euclidean space, the equation describing 
a pion-like Nambu-Goldstone boson in QED in a magnetic field has the form of a 
two-dimensional Schr\"{o}dinger equation \cite{Gusynin:1995gt}:
\begin{equation}
\left[-\frac{\partial^2}{\partial r_1^2}- \frac{\partial^2}{\partial r_2^2} +m^2+V(\mathbf{r})\right]
\psi(\mathbf{r})=0.
\label{eq:schre}
\end{equation}
The function $\psi(\mathbf{r})$ is defined in terms of the Bethe-Salpeter wave function 
$A(p)$ in exactly the same way as in the NJL model, see Eq.~(\ref{wave-function}). 
This time, however, $A(p)$ satisfies a different integral equation, 
\begin{equation}
A(p)=\frac{\alpha}{2\pi^2}\int\frac{d^2k A(k)}{k^2+m^2}
\int\limits^{\infty}_{0} \frac{dx\, \exp(-x\ell^2/2)}{({\bf k-p})^2+x},
\label{eq:Ap}
\end{equation}
where $\ell=1/\sqrt{|e B |}$ is the magnetic length.
Note that, in addition to using the lowest Landau level approximation, we assumed that 
the photon screening effects are negligible. As is easy to check, the explicit form of the 
potential $V(\mathbf{r})$ is given by \cite{Gusynin:1995gt}
\begin{equation}
V(\mathbf{r})=\frac{\alpha}{\pi \ell^2}\exp{\left(\frac{r^2}{2\ell^2}\right)}
\mbox{Ei}\left(-\frac{r^2}{2\ell^2}\right),
\end{equation}
where $r^2=r_1^2+r_2^2$ and $\mbox{Ei}(x)=-\int_{-x}^{\infty}dt \exp(-t)/t$ 
is the integral exponential function \cite{1980tisp.book.....G}. Since $V(\mathbf{r})$ is 
negative, we have a Schr\"{o}dinger equation with an attractive potential, in which 
the parameter $-m^2$ plays the role of the energy $E$. Therefore, the problem 
is again reduced to finding the spectrum of bound states with $E=-m^2<0$. 

It is known that the energy of the lowest level $E(\alpha)$ for the two-dimensional 
Schr\"{o}dinger equation is a nonanalytic function of the coupling constant $\alpha$ 
at $\alpha=0$ \cite{Simon:1976un}. If the potential $V(\mathbf{r})$ were short-range, 
the result would have the form $m^2=-E(\alpha)\propto \exp\left[-1/(C \alpha)\right]$,
 where $C$ is a positive constant \cite{Simon:1976cx,Simon:1976un}. In our case, however, we
have a long-range potential. Indeed, using the asymptotic expansion for $\mbox{Ei}(x)$ 
\cite{1980tisp.book.....G}, we get:
\begin{equation}
V(\mathbf{r}) \simeq -\frac{2\alpha}{\pi}\frac{1}{r^2}, \qquad r\to\infty.
\end{equation}
In order to find an approximate solution for $m^2$, one can use the integral 
equation (\ref{eq:Ap}) at $p=0$. As $\alpha\to 0$, the dominant contribution in the
integral on the right-hand side comes from the infrared region $k^2\lesssim m^2$. 
Therefore, 
\begin{equation}
A(0)\simeq \frac{\alpha}{2\pi^2}A(0)\int\frac{d^2k}{k^2+m^2}
\int\limits^{\infty}_{0} \frac{dx \exp(-y/2)}{l^2k^2+y}\simeq
\frac{\alpha}{4\pi}A(0)\left[\ln\left(\frac{m^2\ell^2}{2}\right)\right]^2,
\end{equation}
which implies that \cite{Gusynin:1995gt}
\begin{equation}
m\propto \sqrt{|eB|}\exp\left(-\sqrt{\frac{\pi}{\alpha}}\right).
\label{eq:m-dyn1}
\end{equation}
A slightly more careful analysis of the integral equation (\ref{eq:Ap}) can be 
made by approximating the interaction kernel so that the exchange momentum 
$(\mathbf{k}-\mathbf{p})^2$ in the denominator is replaced by $\mbox{max}(k^2,p^2)$.
The problem then reduces to an ordinary differential equation with two (infrared 
and ultraviolet) boundary conditions. The approximate analytical solution reveals 
that the lowest energy bound state, which describes the stable vacuum solution 
in quantum field theory, corresponds to the following value of the dynamical mass
\cite{Gusynin:1995gt}:
\begin{equation}
m\simeq C \sqrt{|eB|}\exp\left(-\frac{\pi}{2}\sqrt{\frac{\pi}{2\alpha}}\right) .
\label{eq:m-dyn2}
\end{equation}
Unfortunately, the approximation used in this analysis is not completely reliable. 
There are higher order diagrams that can substantially modify the interaction potential
and, in turn, the result for the dynamical mass. For example, taking into account the 
vacuum polarization effects in the improved rainbow (ladder) approximation, in which 
the free photon propagator is replaced by a screened interaction with the one-loop 
photon self-energy, the result changes. The corrected expression for the mass has 
the same form as in Eq.~(\ref{eq:m-dyn2}), but with $\alpha$ replaced by $\alpha/2$ 
\cite{Gusynin:1995gt}. This is a clear indication that, despite weak coupling, 
there can exist other relevant contributions, coming from higher order diagrams.

A further study showed that, by using a similarity between the magnetic catalysis problem 
in QED and the exactly solvable Schwinger model \cite{Schwinger:1962tp,Frishman:1975}, 
one can find a special nonlocal gauge, in which the leading singularity of the dynamical 
mass can be extracted exactly \cite{Gusynin:1998zq}, 
\begin{equation}
m\simeq \tilde C\sqrt{|eB|} F(\alpha) \exp\left[-\frac{\pi} {\alpha\ln\left(C_1/N\alpha\right)}\right],
\label{eq:m-dyn3}
\end{equation}
where $N$ is the number of fermion flavors, $F(\alpha)\simeq(N\alpha)^{1/3}$, 
$C_1\simeq 1.82\pm 0.06$ and $\tilde C\sim O(1)$. Note that the leading singularity 
in the final expression for the mass is quite different from that in the rainbow 
approximation (\ref{eq:m-dyn2}). 

The magnetic catalysis of chiral symmetry breaking in QED yields a rare example 
of dynamical symmetry breaking in a $(3+1)$-dimensional gauge theory without 
fundamental scalar fields, in which there exists a consistent truncation of the 
Schwinger-Dyson equation.

\subsection{Magnetic catalysis in  QCD}
\label{sec:3.2}
 
Recently there was an increased interest in studies of QCD in a strong magnetic field
\cite{Kabat:2002er,Miransky:2002rp,Agasian:2008zz,Buividovich:2009bh,Mizher:2008hf,
Fraga:2008qn,Fraga:2008um,Fraga:2009vy,Gatto:2010qs,Gatto:2010pt,
Chernodub:2010qx,Fraga:2010qb,Mizher:2010zb,Mizher:2011wd,Frasca:2011zn,
2010PhRvD..82e1501D,Bali:2011uf,Bali:2011qj,Bruckmann:2011zx,Bali:2012zg,Fraga:2012fs}. 
There are several reasons why such investigations may be of interest. Very strong 
magnetic fields are known to have existed in the Early Universe \cite{Vachaspati:1991nm,
Enqvist:1993kf,Cheng:1994yr,Baym:1995fk,Grasso:2000wj} and are expected 
to be generated in relativistic heavy ion collisions \cite{Kharzeev:2007jp,Skokov:2009qp}. 
Since the chiral symmetry plays a profound role in QCD, it is interesting to study also the 
role of magnetic catalysis in this theory \cite{Miransky:2002rp,Kabat:2002er}.

Because of the property of asymptotic freedom, one can argue that the dynamics 
underlying magnetic catalysis in QCD is, at least in principle, weakly coupled 
at sufficiently large magnetic fields \cite{Kabat:2002er}. This fact can be used to 
justify a consistent truncation of the Schwinger-Dyson equation, resembling that 
in QED, which we discussed in the preceding section. 

Let us start by introducing a QCD like theory with $N_u$ up flavors of quarks having 
electric charges $2e/3$ and $N_d$ down flavors of quarks having electric charges 
$-e/3$. (The total number of flavors is $N_f=N_u+N_d$.) It is important to distinguish 
the up and down types of quarks because the chiral symmetry subgroup that mixes 
them is explicitly broken by the external magnetic field. Taking this into account, we 
find that the model is invariant under the $SU(N_{u})_{L}\times SU(N_{u})_{R} 
\times SU(N_{d})_{L}\times SU(N_{d})_{R}\times U^{(-)}(1)_{A}$ chiral symmetry. 
The anomaly free subgroup $U^{(-)}(1)_{A}$ is connected with the conserved 
current which is the difference of the $U^{(d)}(1)_{A}$ and $U^{(u)}(1)_{A}$ currents. 
[The $U^{(-)}(1)_{A}$ symmetry is of course absent when either $N_d$ or $N_u$ 
equals zero.] A dynamical generation of quark masses spontaneously breaks the 
chiral symmetry down to $SU(N_{u})_{V}\times SU(N_{d})_{V}$ and gives rise to
$N_{u}^{2}+N_{d}^{2}-1$ massless Nambu-Goldstone bosons in the 
low-energy spectrum.

Just like in QED, the vacuum polarization effects play a very important role in 
QCD in the presence of a strong magnetic field. By properly modifying the known 
result from the Abelian gauge theory \cite{1976PhLA...56..151L,Dittrich:1985yb,
Calucci:1993fi} to the case of QCD, we find that the gluon 
polarization tensor has the following behavior:
\begin{eqnarray}
\Pi^{AB,\mu\nu} &\simeq& \frac{\alpha_{s}}{6\pi}
\delta^{AB} \left(k_{\parallel}^{\mu}
k_{\parallel}^{\nu}-k_{\parallel}^{2}g_{\parallel}^{\mu\nu}\right)
\sum_{q=1}^{N_{f}}\frac{|e_{q}B|}{m^{2}_{q}},
\quad \mbox{for} \quad |k_{\parallel}^2| \ll m_{q}^2,
\label{Pi-IR}\\
\Pi^{AB,\mu\nu} & \simeq& -\frac{\alpha_{s}}{\pi}
\delta^{AB} \left(k_{\parallel}^{\mu}
k_{\parallel}^{\nu}-k_{\parallel}^{2}g_{\parallel}^{\mu\nu}\right)
\sum_{q=1}^{N_{f}}\frac{|e_{q}B|}{{k_{\parallel}^2}}, 
\quad \mbox{for} \quad m_{q}^2 \ll |k_{\parallel}^2|\ll |eB|,
\label{Pi-UV}
\end{eqnarray}
where $k_{\parallel}^{\mu}\equiv g_{\parallel}^{\mu\nu} k_{\nu}$ and 
$g_{\parallel}^{\mu\nu}\equiv \mbox{diag}(1,0,0,-1)$ is the projector onto the 
longitudinal subspace. Notice that quarks in a strong magnetic field do not 
couple to the transverse subspace spanned by $g_{\perp}^{\mu\nu}\equiv 
g^{\mu\nu} -g_{\parallel}^{\mu\nu} =\mbox{diag}(0,-1,-1,0)$ and 
$k_{\perp}^{\mu}\equiv g_{\perp}^{\mu\nu} k_{\nu}$. This is connected with
the dominant role of the lowest Landau level, in which quarks are polarized along the
magnetic field.

The expressions (\ref{Pi-IR}) and (\ref{Pi-UV}) coincide with those for the 
polarization operator in the massive Schwinger model \cite{Schwinger:1962tp} 
if the parameter $\alpha_{s} |e_{q}B|/2$ here is replaced by the dimensional 
coupling $\alpha_{1}$ of $(1+1)$-dimensional QED. In particular, 
Eq.~(\ref{Pi-UV}) implies that there is a massive gluon resonance with 
the mass given by
\begin{equation}
M_{g}^2= \sum_{q=1}^{N_{f}}\frac{\alpha_{s}}{\pi}|e_{q}B|=
(2N_{u}+N_{d}) \frac{\alpha_{s}}{3\pi}|eB|.
\label{M_g}
\end{equation}
This is reminiscent of the pseudo-Higgs effect in the (1+1)-dimensional
massive QED. It is not the genuine Higgs effect because there is no
complete screening of gluons in the far infrared region with
$|k_{\parallel}^2|\ll m_{q}^2$, see Eq.~(\ref{Pi-IR}). Nevertheless, the 
pseudo-Higgs effect is manifested in creating a massive resonance and 
this resonance provides the dominant force leading to chiral symmetry 
breaking.

In the end, the dynamics in QCD in a strong magnetic field appears to be essentially 
the same as in QED, except for purely kinematic changes. After expressing 
the magnetic field in terms of the running coupling $\alpha_{s}$ at the scale 
$\sqrt{|eB|}$ using
\begin{equation}
\frac{1}{\alpha_{s}} \simeq b\ln\frac{|eB|}{\Lambda_{QCD}^2},
\quad \mbox{ where} \quad
b=\frac{11 N_c -2 N_f}{12\pi},
\label{coupling}
\end{equation}
we obtain the result for the dynamical mass in the following form \cite{Miransky:2002rp}:
\begin{equation}
m_{q}^2 \simeq 2C_{1}\left|\frac{e_{q}}{e}\right| \Lambda_{QCD}^2 
\left(c_{q}\alpha_{s}\right)^{2/3}
\exp\left[\frac{1}{b\alpha_{s}}-\frac{4N_{c}\pi}{\alpha_{s} 
(N_{c}^{2}-1) \ln(C_{2}/c_{q}\alpha_{s})}\right],
\label{gap-vs-alpha}
\end{equation}
where $e_{q}$ is the electric charge of the $q$-th quark and $N_{c}$
is the number of colors. The numerical factors $C_1$ and $C_2$ are 
of order $1$, and the value of $c_{q}$ is given by
\begin{equation}
c_{q} = \frac{1}{6\pi}(2N_{u}+N_{d})\left|\frac{e}{e_{q}}\right| .
\end{equation}
Because of the difference in electric charges, the dynamical mass of the up-type quarks 
is considerably larger than that of the down-type quarks. 

It is interesting to point that the dynamical quark masses in a wide range 
of strong magnetic fields, $\Lambda_{QCD}^{2}\ll |eB| \lesssim (10 \mbox{ TeV})^{2}$, 
remain much smaller than the dynamical (constituent) masses of quarks $m_q^{(0)} \simeq 
300\,\mbox{MeV}$ in vacuum QCD {\em without} a magnetic field. This may suggest that 
QCD can have an intermediate regime, in which the magnetic field is strong enough to 
provide a gluon screening to interfere with the vacuum pairing dynamics \cite{Miransky:2002rp,Galilo:2011nh}, 
but not sufficiently strong to produce large dynamical masses through magnetic catalysis. 
In this intermediate regime, the dynamical mass and the associated chiral condensate could 
be {\em decreasing} with the magnetic field. The corresponding regime may start already 
at magnetic fields as low as $10^{19}\,\mbox{G}$, when the gluon mass $M_g$, given
by Eq.~(\ref{M_g}), becomes comparable to $\Lambda_{\rm QCD}$. (For the estimate,
we assumed that the value of the coupling constant is of order $1$ at the QCD energy scale.)

\subsection{Magnetic Catalysis in Graphene}
\label{sec:3.3}

In this section, we briefly discuss the application of the magnetic catalysis ideas to 
graphene in the regime of the quantum Hall effect. 

Graphene is a single atomic layer of graphite \cite{2004Sci...306..666N} that has 
many interesting properties and promises widespread applications (for reviews,
see Refs.~\cite{Gusynin:2007ix,2009RvMP...81..109C,Abergel:2010}). The 
uniqueness of graphene is largely due to its unusual band structure with two  
Dirac points at the corners of the Brillouin zone. Its low-energy excitations are 
described by massless Dirac fermions \cite{Semenoff:1984dq}. Because of
a relatively small Fermi velocity of quasiparticles, $v_F\approx c/300$, the 
effecting coupling constant for the Coulomb interaction in graphene, 
$\alpha \equiv {e^2}/(\epsilon_0 v_F)$, is about $300$ times larger 
than the fine structure constant in QED, $e^2/(\epsilon_0 c)\approx 1/137$.

When graphene is placed in a perpendicular magnetic field, it reveals an anomalous 
quantum Hall effect \cite{Novoselov:2005kj,Zhang:2005zz}, exactly as predicted in theory 
\cite{Zheng:2002zz,Gusynin:2005pk,Peres:2006zz}. The anomalous plateaus 
in the Hall conductivity are observed at the filling factors $\nu = \pm 4(n + 1/2)$, 
where $n=0,1,2,\ldots$ is the Landau level index. The factor $4$ in the filling 
factor is due to a fourfold (spin and valley) degeneracy of each Landau level. 
As for the half-integer shift in the filling factor, it is directly connected with the 
Dirac nature of quasiparticles \cite{Semenoff:1984dq,Haldane:1988zza,
Khveshchenko:2001zza,Khveshchenko:2001zz,Gorbar:2002iw}.

It was observed experimentally \cite{2006PhRvL..96m6806Z,2007PhRvL..99j6802J,
2009Natur.462..192D,2009Natur.462..196B} that there appear additional plateaus 
in the Hall conductivity when graphene is placed in a very strong magnetic field. 
The new plateaus can be interpreted as the result of lifting the fourfold degeneracy 
of the Landau levels. In the case of the lowest Landau level, in particular, some of the degeneracy, 
i.e., between the particle and hole states, can be removed when there is a dynamical 
generation of a Dirac mass. Considering the possibility of magnetic catalysis, 
such an outcome seems almost unavoidable \cite{Khveshchenko:2001zza,Khveshchenko:2001zz,
Gorbar:2002iw,Gusynin:2006gn,Gorbar:2008hu,Herbut:2008ui,Semenoff:2011ya,
Gorbar:2011kc}. 

The low-energy quasiparticle excitations in graphene are conveniently described in terms 
of four-component Dirac spinors $\Psi_s^T = \left( \psi_{KAs},\psi_{KBs},\psi_{K^\prime Bs},
\psi_{K^\prime As}\right)$, introduced for each spin state $s=\uparrow,\downarrow$. Note that 
the components of $\Psi_s$ are the Bloch states from two sublattices ($A,B$) of the 
graphene hexagonal lattice and two valleys ($K,K^\prime$) at the opposite corners of the 
Brillouin zone. The approximate low-energy 
Hamiltonian, including the kinetic and Coulomb interaction terms, is given by
\begin{eqnarray}
 H &=& v_F\sum_{s}\int d^2\mathbf{r}\,\overline{\Psi}_s\left(\gamma^1\pi_x+
\gamma^2\pi_y\right)\Psi_s \nonumber\\
&&+\frac{1}{2} \sum_{s,s^\prime} \int d^2\mathbf{r}d^2\mathbf{r}^\prime 
{\Psi}^{\dagger}_{s}(\mathbf{r}) \Psi_s(\mathbf{r})
U_{C}(\mathbf{r}-\mathbf{r}^\prime)  
{\Psi}^{\dagger}_{s^\prime}(\mathbf{r}^\prime) \Psi_{s^\prime}(\mathbf{r}^\prime),
\label{graphene-hamiltonian}
\end{eqnarray}
where $U_{C}(\mathbf{r})$ is the Coulomb potential, which takes into account the polarization 
effects in a magnetic field \cite{Gorbar:2002iw,Gorbar:2011kc}. Note that the two electron 
spins ($s=\uparrow,\downarrow$) in graphene give rise to two independent species of 
Dirac fermions. As a result, the Hamiltonian possesses an approximate $U(4)$ symmetry 
\cite{Gorbar:2002iw}, which is a generalization of the $U(2)$ flavor symmetry discussed in the 
case of the one-species model in Sec.~\ref{mag-cat-2D}. The 16 generators of the extended 
$U(4)$ flavor symmetry are obtained by a direct product of the 4 generators of the $U(2)$ group
acting in the valley space $(K,K^\prime)$, and the 4 generators of the $U(2)$ spin symmetry. 

The $U(4)$ symmetry is preserved even when the electron chemical potential term, 
$-\mu \Psi^{\dagger}\Psi$, is added. The inclusion of the Zeeman term, 
which distinguishes the electron states with opposite spins, breaks the symmetry 
down to the $U_{\uparrow}(2) \times U_{\downarrow}(2)$ subgroup. The explicit form 
of the Zeeman term is given by $\mu_BB\Psi^{\dagger} \sigma_3 \Psi$, where $B$ is 
the magnetic field, $\mu_B=e\hbar/(2mc)$ is the Bohr magneton, and $\sigma_3$ is 
the third Pauli matrix in spin space. An interesting thing is that this explicit symmetry breaking is 
a small effect even in very strong magnetic fields. To see this, we can compare the 
Zeeman energy $\varepsilon_Z$ with the Landau energy $\varepsilon_\ell$,
\begin{eqnarray}
\varepsilon_Z  &=& \mu_B B =5.8\times 10^{-2}B[\mbox{T}]~\mbox{meV},\\
\varepsilon_\ell &=& \sqrt{\hbar v_F^2|eB|/c} =26 \sqrt{B[\mbox{T}]}~\mbox{meV}.
\end{eqnarray}
Therefore, the Zeeman energy is less then a few percent of the Landau energy 
even for the largest (continuous) magnetic fields created in a laboratory, 
$B\lesssim 50\,\mbox{T}$.

Because of the large flavor symmetry, there are many potential ways how it can be 
broken \cite{Gorbar:2008hu,Gorbar:2011kc}. Here we mention only the possibilities 
that are connected to the magnetic catalysis scenario at zero filling $\nu=0$ 
(i.e., the lowest Landau level is half-filled). 

We will allow independent symmetry breaking condensates for fermions with opposite 
spins. Also, in addition to the usual $\langle \bar{\Psi}_s \Psi_s \rangle$ condensates 
(no sum over the repeated spin indices here), we introduce the time reversal odd ones, 
$\langle \bar{\Psi}_s \gamma^3 \gamma^5 \Psi_s \rangle$ \cite{Gorbar:2008hu,Gorbar:2011kc}.
While the former will give rise to Dirac masses $m_{s}$ ($s=\uparrow,\downarrow$) 
in the low-energy theory, the latter will result in the Haldane masses $\Delta_{s}$ 
($s=\uparrow,\downarrow$) \cite{Haldane:1988zza}.

In the ground state, one can also have additional condensates, $\langle \Psi^{\dagger}\sigma^3 \Psi \rangle$
and $\langle \Psi^{\dagger}\gamma^3\gamma^5P_{s}\Psi \rangle$, associated with nonzero 
spin and pseudo-spin (valley) densities. To capture this possibility in the variational ansatz,
one needs to include a spin chemical potential $\mu_3$ and two pseudo-spin chemical potentials
$\tilde{\mu}_s$ ($s=\uparrow,\downarrow$). Thus, the general structure of the (inverse) full fermion 
propagator for quasiparticles of a fixed spin has the following form:
\begin{equation}
S_s^{-1}(\omega;\mathbf{r},\mathbf{r}^\prime)= - i\left[\gamma^0 \omega
- v_F  (\bm{\pi}\cdot\bm{\gamma})
+\hat{\Sigma}_s^{+} \right]\delta^{2}(\mathbf{r}-\mathbf{r}^\prime),
\label{inversefull}
\end{equation}
where the generalized self-energy operator $\hat{\Sigma}^{+}$ is given by
\begin{equation}
\hat{\Sigma}^{+} = -m_s + \gamma^0\mu_s
+  i s_{\perp}\gamma^1\gamma^2 \tilde\mu_s
+  i s_{\perp}\gamma^0 \gamma^1\gamma^2 \Delta_s.
\label{def-S+TXT}
\end{equation}
Functions $m_s$, $\mu_s$, $\tilde{\mu}_s$, and $\Delta_s$ on the right-hand side 
depend on the operator valued argument 
$(\bm{\pi}\cdot\bm{\gamma})^{2}\ell^2$, whose eigenvalues are nonpositive even 
integers: $-2n$, where $n=0,1,2,\ldots$. Therefore, in the Landau level representation, 
$m_s$, $\mu_s$, $\tilde{\mu}_s$, and $\Delta_s$ will get an additional Landau  
index $n$ dependence: $m_{n,s}$, $\mu_{n,s}$, $\tilde{\mu}_{n,s}$, and $\Delta_{n,s}$.

The Schwinger--Dyson equation for the full fermion propagator takes the form
\begin{equation}
S^{-1}(t-t^\prime;\mathbf{r},\mathbf{r}^{\prime}) = S_0^{-1}(t-t^\prime;\mathbf{r},\mathbf{r}^{\prime})
+ e^2\gamma^0\,S(t-t^\prime;\mathbf{r},\mathbf{r}^{\prime})\gamma^0 D(t^\prime-t;\mathbf{r}^{\prime}-\mathbf{r}),
\label{SD}
\end{equation}
where $D(t;\mathbf{r})$ is the photon propagator mediating the Coulomb interaction. The
latter is approximately instantaneous because the quasiparticle velocities are much 
smaller than the speed of light. In momentum space, the photon propagator takes the following form:
\begin{equation}
D(\omega,k)\approx D(0,k)=\frac{i}{\epsilon_0 [k+\Pi(0,k)]} ,
\end{equation}
where $\Pi(0,k)$ is the static polarization function and $\epsilon_0$ is a dielectric constant.

It should be noted that, in the coordinate-space representation, both the fermion propagator 
and its inverse contain exactly the same Schwinger phase, see Eq.~(\ref{schwinger-phase}). 
After omitting such a (nonzero) phase on both sides of Eq.~(\ref{SD}) and performing the 
Fourier transform with respect to the time variable, we will arrive at the following equation 
for the translationally invariant part of the fermion propagator \cite{Gorbar:2011kc}:
\begin{equation}
\tilde{S}^{-1}(\omega;\mathbf{r}) = \tilde{S}_0^{-1}(\omega;\mathbf{r})
+i\frac{e^2}{\epsilon_0} \int_{-\infty}^{\infty} \frac{d\Omega}{2\pi} \int_{0}^{\infty} \frac{dk}{2\pi}
\frac{k J_{0}(kr)}{k+\Pi(0,k)} \gamma^0\,\tilde{S}(\Omega;\mathbf{r})\gamma^0\,.
\label{SD-omega-TXT}
\end{equation}
In the Landau level representation, this equation is equivalent to a coupled set of $4\times 2\times 
n_{\rm max}$ equations, where we counted 4 parameters ($m$, $\mu$, $\tilde{\mu}$, and $\Delta$), 
2 spins ($s=\uparrow,\downarrow$), and $n_{\rm max}\simeq \left[\Lambda^2/(2|eB|)\right]$ Landau 
levels below the ultraviolet energy cutoff $\Lambda$, where the low-energy theory is valid. 

The explicit form of the gap equations can be found elsewhere \cite{Gorbar:2011kc}. The corresponding 
set of equations can be solved by making use of numerical methods. Here, instead, we will discuss only 
some general features of the solutions in the lowest Landau level approximation, which can be obtained with analytical methods. 

Let us start by considering the solutions to the gap equations for quasiparticles of a fixed spin.
In the lowest Landau level approximation, there are two independent gap equations, i.e.,
\begin{eqnarray}
\mu_{\rm eff}-\mu&=& \frac{\alpha \varepsilon_\ell }{2} {\cal K}_{0}   \left[
n_{F}\left(m_{\rm eff}-\mu_{\rm eff}\right)-n_{F}
\left(m_{\rm eff}+\mu_{\rm eff}\right)
\right],\label{eq-mu0-eff-0}\\
m_{\rm eff}&=& \frac{\alpha \varepsilon_\ell }{2} {\cal K}_{0}     \left[
1-n_{F}\left(m_{\rm eff}-\mu_{\rm eff}\right)-n_{F}
\left(m_{\rm eff}+\mu_{\rm eff}\right)
\right],
\label{eq-m0-eff-0}
\end{eqnarray}
where $\alpha\equiv {e^2}/(\epsilon_0 v_F)\approx 2.2/\epsilon_0$ is the coupling constant,  
$n_{F}(x)\equiv 1/\left(e^{x/T}+1\right)$ is the Fermi distribution function, and 
${\cal K}_{0} $ is the interaction kernel due to the Coulomb interaction in the 
lowest Landau level approximation. In the above equations, 
we used the shorthand notation $\mu_{\rm eff} = \mu-\Delta$ and $m_{\rm eff} =m -\tilde{\mu}$ 
for the two independent  combination of parameters that determine the spectrum of  
the lowest Landau level quasiparticles,
\begin{equation}
\omega_{-} =-\mu_{\rm eff} - m_{\rm eff}  ,
\qquad\mbox{and}\qquad
\omega_{+} = -\mu_{\rm eff} + m_{\rm eff}  .
\end{equation}
At zero temperature, the gap equations reduce down to
\begin{eqnarray}
\mu_{\rm eff}  &= & \mu+  \frac{\alpha\varepsilon_\ell}{4\sqrt{2\pi}}\mbox{sign}(\mu_{\rm eff} )
\theta(|\mu_{\rm eff} |-|m_{\rm eff} |),\\
|m_{\rm eff} | &=&  \frac{\alpha\varepsilon_\ell}{4\sqrt{2\pi}}
\theta(|m_{\rm eff} |-|\mu_{\rm eff} |).
\end{eqnarray}
Here we used the value for the interaction kernel ${\cal K}_{0}  = 1/(2\sqrt{2\pi})$, which is 
obtained in the approximation with screening effects neglected \cite{Gorbar:2011kc}.
One of the solutions to this set of equations has a nonzero dynamical Dirac mass
($m \propto \alpha\varepsilon_\ell$), i.e.,
\begin{equation}
|m_{\rm eff} | =   \frac{\alpha\varepsilon_\ell}{4\sqrt{2\pi}} ,
\qquad 
\Delta  = 0, \qquad 
- \frac{\alpha\varepsilon_\ell}{4\sqrt{2\pi}}  < \mu < \frac{\alpha\varepsilon_\ell}{4\sqrt{2\pi}} .
\label{sol-Dirac}  
\end{equation}
The other two solutions have nonzero Haldane masses ($\Delta \propto \alpha\varepsilon_\ell$), i.e.,
\begin{eqnarray}
&&
m_{\rm eff}  = 0, \qquad 
\Delta  = \frac{\alpha\varepsilon_\ell}{4\sqrt{2\pi}} , \qquad 
-\infty<\mu< \frac{\alpha\varepsilon_\ell}{4\sqrt{2\pi}} ,
\label{sol-Haldane-1}   \\
&&
m_{\rm eff}  = 0,\qquad
\Delta  = -\frac{\alpha\varepsilon_\ell}{4\sqrt{2\pi}} , \qquad 
-\frac{\alpha\varepsilon_\ell}{4\sqrt{2\pi}} <\mu<\infty .
\label{sol-Haldane-2}  
\end{eqnarray}
In both types of solutions, the values of the masses are proportional to a power of the 
coupling constant $\alpha$, as expected from the dimensional reduction
\cite{Landau:101811,Simon:1976cx,Simon:1976un}. 

In order to determine the ground state in graphene when both spin states are accounted for, 
one has to find among many possible solutions the one with the lowest free energy. 
In the approximation used here, the ground state solution at $\nu=0$ filling (i.e., an analog 
of the vacuum state in particle physics) corresponds to a spin-singlet state with equal in 
magnitude, but opposite in sign Haldane masses for the two spin states \cite{Gorbar:2011kc}: 
$\Delta_\uparrow=-\Delta_\downarrow$, i.e., a mixture of the two solutions in 
Eqs.~(\ref{sol-Haldane-1}) and (\ref{sol-Haldane-2}). 

The symmetry of the corresponding ground state is $U_{\uparrow}(2) \times U_{\downarrow}(2)$, 
but with the Zeeman energy splitting dynamically enhanced by the nonzero Haldane masses. 
The quasiparticle energies of the dynamically modified lowest Landau level are \cite{Gorbar:2011kc}
\begin{eqnarray}
\omega_{\uparrow} &=& -\mu+ \varepsilon_Z +|\Delta_{\uparrow}| >0  ,\qquad (\times 2),\\
\omega_{\downarrow} &=& -\mu-  \varepsilon_Z -|\Delta_{\downarrow}| <0,\qquad (\times 2),
\end{eqnarray}
which show that the original fourfold degeneracy is indeed partially lifted.

\section{Concluding Remarks}
\label{sec:4}

We hope that this review of magnetic catalysis is sufficient to convey the main 
idea of the phenomenon in terms of simple and rather general physics concepts. 
From the outset, this review was never meant to be comprehensive. Here we
concentrated only on the bare minimum needed to understand the phenomenon 
as a consequence of the underlying dimensional reduction of the fermion-antifermion pairing 
in a magnetic field \cite{Gusynin:1994re,Gusynin:1994va,Gusynin:1994xp,Gusynin:1995gt}. 
For further reading and for deeper insights into various aspects of the magnetic 
catalysis, it is suggested that the reader refers to the original literature on the topic. 

Over nearly 20 years of research, there has been a lot of progress made in our 
understanding of magnetic catalysis. A rather long list of research papers 
at the end of this review is a pretty objective proof of that. At present, it is evident 
that the key features of the underlying physics are well established and understood. 
At the same time, it is also evident that there are still many theoretical questions about the 
applications of magnetic catalysis under various conditions, where factors other 
than the magnetic field may also play a substantial role. 

One prominent example is the dynamics of chiral symmetry breaking in QCD in a 
magnetic field. Because of a poorly understood interplay between the dynamics 
responsible for the quark (de-)confinement on the one hand and the magnetic catalysis 
on the other, there are a lot of uncertainties about the precise role of the magnetic field in 
this case \cite{Kabat:2002er,Miransky:2002rp,Agasian:2008zz,Buividovich:2009bh,Mizher:2008hf,
Fraga:2008qn,Fraga:2008um,Fraga:2009vy,Gatto:2010qs,Gatto:2010pt,
Chernodub:2010qx,Fraga:2010qb,Mizher:2010zb,Mizher:2011wd,Frasca:2011zn,
2010PhRvD..82e1501D,Bali:2011uf,Bali:2011qj,Bruckmann:2011zx,Bali:2012zg,Fraga:2012fs}. 
One can even suggest that there exists an intermediate regime in QCD, starting 
at magnetic fields of order $B\simeq 10^{19}\,\mbox{G}$ or so, in which the 
magnetic field is sufficiently strong to provide a gluon screening 
\cite{Galilo:2011nh} and, thus, suppress the vacuum chiral 
condensate, but still is not strong enough to produce equally large 
quark masses through magnetic catalysis \cite{Miransky:2002rp}. 
At finite temperature, further complications could appear because of the interplay 
of the magnetic field and the temperature in gluon screening \cite{Bali:2011qj}. 
All in all, it is obvious that there are many research directions remaining to be 
pursued in the future. 

As we argued in Sec.~\ref{sec:3.3}, magnetic catalysis may play a profound role 
in the quantum Hall effect in monolayer graphene. It appears, however,
that an interesting variation of magnetic catalysis can be also realized in bilayer 
graphene \cite{2010JETPL..91..314G,2010PhRvB..81o5451G,Gorbar:2011ce,
Gorbar:2012jc}. In essence, it is a {\em nonrelativistic} analog of the magnetic 
catalysis. This fact alone is of interest because of a large diversity of solid 
state physics systems and the relative ease of their studies in table-top 
experiments.

Finally, one should keep in mind that the fundamental studies of gauge field 
theories, which are known to have an extremely rich and complicated dynamics, 
is of general interest even in the regimes that are not readily accessible in current 
experiments. Such studies usually provide invaluable information about the complicated 
theories in the regimes that are under theoretical control. This often allows one 
to understand better the structure of the theory and even predict 
its testable limitations. In the case of QCD in a magnetic field, e.g., we may
gain not only a better understand of the fundamental properties, but also 
get an insight into the physics in the Early Universe and in heavy ion 
collisions.

\begin{acknowledgement}
The author thanks E.V.~Gorbar, V.P.~Gusynin and V.A.~Miransky for reading the 
early version of the review and offering many useful comments.
This work was supported in part by the U.S. National Science Foundation 
under Grant No. PHY-0969844.
\end{acknowledgement}

\appendix 
\section*{Appendix: Fermion propagator in a magnetic field}
\addcontentsline{toc}{section}{Appendix}
\label{sec:Appendix}

Let us start from the discussion of the Dirac fermion propagator in a magnetic field in 
$3+1$ dimensions. It is formally defined by the following expression:
\begin{equation}
S(x,x^\prime)
= i \left[ i \gamma^0 \partial_t - (\bm{\pi}_{\perp}\cdot\bm{\gamma}_{\perp})-\pi^{3}\gamma^3-m\right]^{-1} 
\delta^{4}(x-x^\prime),
\label{prop-S1-app}
\end{equation}
where $x\equiv (x^0,x^1,x^2,x^3) =(t,\mathbf{r})$. By definition, the spatial components of the 
canonical momenta are $\pi^{i} \equiv -i \partial_i - e A^i$, where $i=1,2,3$. (The perpendicular 
components are $i=1,2$.) Here we 
assume that $e$ is the fermion electric charge (i.e., one should take $e<0$ in the case of the 
electron)  and use the Landau gauge $\mathbf{A}= (0, B x^1,0)$, where $B$ is the magnetic 
field pointing in the $x^3$-direction. By definition, the components of the usual three-dimensional 
vectors $\mathbf{A}$ (vector potential) and $\mathbf{r}$ (position vector) are identified with
the {\em contravariant} components $A^i$ and $x^i$, respectively.

In the Landau gauge used, it is convenient to perform a Fourier transform in the time ($t-t^\prime$) and 
the longitudinal ($x^3-x^{\prime\, 3}$) coordinates. Then, we obtain
\begin{eqnarray}
S(\omega,p^3;\mathbf{r}_{\perp},\mathbf{r}_{\perp}^\prime)
&=& i \left[\gamma^0 \omega -(\bm{\pi}_{\perp}\cdot\bm{\gamma}_{\perp}) -\gamma^3 p^3 -m\right]^{-1} 
\delta^{2}(\mathbf{r}_{\perp}-\mathbf{r}_{\perp}^\prime) \nonumber\\
&=&i \left[\gamma^0 \omega -(\bm{\pi}_{\perp}\cdot\bm{\gamma}_{\perp}) -\gamma^3 p^3 +m\right]\nonumber\\
&& \times \left[\omega^2 -\bm{\pi}_{\perp}^2 + i eB\gamma^1\gamma^2 -(p^3)^2 -m^2 \right]^{-1}
\delta^{2}(\mathbf{r}_{\perp}-\mathbf{r}_{\perp}^\prime) ,
\label{prop-S1-omega}
\end{eqnarray}
where $\mathbf{r}_{\perp}$ is the position vector in the plane perpendicular to the magnetic field.

In order to obtain a Landau level representation for the propagator (\ref{prop-S1-omega}), it is 
convenient to utilize the complete set of eigenstates of the operator $\bm{\pi}_{\perp}^{2}$. 
This operator has the eigenvalues $(2k+1)|eB|$, where $k=0,1,2,\dots$ is the quantum number 
associated with the orbital motion in the perpendicular plane. The corresponding normalized 
wave functions read
\begin{equation}
\psi_{k p_2}(\mathbf{r}_{\perp})=\frac{1}{\sqrt{2\pi \ell}}\frac{1}{\sqrt{2^k k!\sqrt{\pi}}}
H_k\left(\frac{x^1}{\ell}+p_2\ell\right)e^{-\frac{1}{2\ell^2}(x^1+p_2\ell^2)^2} e^{-i s_{\perp}x^2 p_2},
\label{wave-fun-orbital}
\end{equation}
where $H_{k}(z)$ are the Hermite polynomials \cite{1980tisp.book.....G}, 
$\ell=1/\sqrt{|e B |}$ is the magnetic length, and $s_{\perp} \equiv \mbox{sign}(eB)$. 
The wave functions satisfy the conditions of normalizability and completeness,
\begin{eqnarray}
\int d^{2} \mathbf{r}_{\perp}\,\psi^{*}_{k p_2}(\mathbf{r}_{\perp})\psi_{k^{\prime}p_2^{\prime}}(\mathbf{r}_{\perp}) 
&=& \delta_{kk^{\prime}} \delta(p_2-p_2^{\prime}), \\
\int\limits_{-\infty}^{\infty} dp_2\, \sum\limits_{k=0}^{\infty}\,
\psi_{kp_2}(\mathbf{r}_{\perp}) \psi^{*}_{kp_2}(\mathbf{r}_{\perp}^{\prime}) 
&=& \delta^{2}(\mathbf{r}_{\perp}-\mathbf{r}_{\perp}^{\prime}),
\label{completeness}
\end{eqnarray}
respectively. 

By making use of the spectral expansion of the $\delta$-function in Eq.~(\ref{completeness}),
as well as the following identities:
\begin{eqnarray}
(\bm{\pi}_{\perp}\cdot\bm{\gamma}_{\perp})\psi_{k p_2} 
&=&  \frac{i}{\ell}\gamma^1
\left[\sqrt{2(k+1)}\, \psi_{k+1,p_2} {\cal P}_{-}
-\sqrt{2k}\, \psi_{k-1,p_2} {\cal P}_{+} \right] ,\\
\bm{\pi}_{\perp}^2 \psi_{k p_2} &=& \frac{2k+1}{\ell^2}\psi_{k p_2} ,
\end{eqnarray} 
with ${\cal P}_{\pm} = \frac{1}{2}\left(1\pm i s_{\perp} \gamma^1\gamma^2\right)$
being the spin projectors onto the direction of the magnetic field,
we can rewrite the propagator in Eq.~(\ref{prop-S1-omega}) as follows:
\begin{eqnarray}
S(\omega,p^3;\mathbf{r}_{\perp},\mathbf{r}_{\perp}^\prime)
&=&\int\limits_{-\infty}^{\infty} dp_2\, \sum\limits_{k=0}^{\infty}\,
i \left[\gamma^0 \omega -(\bm{\pi}_{\perp}\cdot\bm{\gamma}_{\perp}) -\gamma^3 p^3 +m\right]
\Big[\omega^2 -(p^3)^2
\nonumber\\
&& -(2k+1)|eB|+i eB\gamma^1\gamma^2 -m^2 \Big]^{-1}
\psi_{kp_2}(\mathbf{r}_{\perp}) \psi^{*}_{kp_2}(\mathbf{r}_{\perp}^{\prime}) 
\nonumber\\
&=& e^{i\Phi(\mathbf{r}_{\perp},\mathbf{r}_{\perp}^{\prime})}
\tilde{S}(\omega,p^3;\mathbf{r}_{\perp}-\mathbf{r}_{\perp}^\prime) .
\label{prop-S2}
\end{eqnarray}
The Schwinger phase is given by 
\begin{equation}
\Phi(\mathbf{r}_{\perp},\mathbf{r}_{\perp}^{\prime}) = s_{\perp} \frac{(x^1+x^{\prime\, 1})(x^2-x^{\prime\, 2})}{2\ell^2},
\label{schwinger-phase}
\end{equation}
and the translationary invariant part of the propagator reads
\begin{eqnarray}
\tilde{S}(\omega,p^3;\mathbf{r}_{\perp}-\mathbf{r}_{\perp}^\prime)
&=& i \frac{e^{-\xi/2}}{2\pi\ell^2} \sum_{n=0}^{\infty} 
\frac{F_n\left(\omega,p^3;\mathbf{r}_{\perp}-\mathbf{r}_{\perp}^\prime\right)}
{\omega^2-2n|eB| -(p^3)^2-m^2} ,
\label{prop-S2-trans}\\
F_n\left(\omega,p^3;\mathbf{r}_{\perp}-\mathbf{r}_{\perp}^\prime\right)&=&
\left(\gamma^0 \omega -\gamma^3 p^3 +m\right) 
\left[ L_{n}(\xi) {\cal P}_{+}+ L_{n-1}(\xi){\cal P}_{-}\right]  \nonumber\\
&&-\frac{i}{\ell^2} \, \bm{\gamma}_{\perp}\cdot(\mathbf{r}_\perp-\mathbf{r}_{\perp}^{\prime})
L_{n-1}^1(\xi),
\label{prop-S2-F-n} 
\end{eqnarray}
where we used the short-hand notation 
\begin{equation}
\xi = \frac{(\mathbf{r}_{\perp}-\mathbf{r}_{\perp}^{\prime})^2}{2\ell^2}.
\end{equation}
In order to integrate over the quantum number $p_2$ in Eq.~(\ref{prop-S2}), we took 
into account the following table integral \cite{1980tisp.book.....G}: 
\begin{equation}
\int\limits_{-\infty}^\infty\,e^{-x^2}H_m(x+y)H_n(x+z)dx
=2^n\pi^{1/2}m!z^{n-m}L_m^{n-m}(-2yz),
 \label{7.378}
\end{equation}
which is valid when $m\le n$. Here $L^{\alpha}_n$ are the generalized Laguerre polynomials, and
$L_n \equiv L^{0}_n$.

Here a short remark is in order regarding the general structure of the Dirac propagator in 
a magnetic field. It is not a translationally invariant function, but has the form of a product 
of the Schwinger phase factor $e^{i\Phi(\mathbf{r}_{\perp},\mathbf{r}_{\perp}^{\prime})}$ 
and a translationally invariant part. The Schwinger phase spoils the translational invariance.
From a physics viewpoint, this reflects a simple fact that the fermion momenta in the two 
spatial directions perpendicular to the field are not conserved quantum numbers. 

The Fourier transform of the translationary invariant part of the propagator (\ref{prop-S2-trans})
reads
\begin{equation}
\tilde{S}(\omega,p^3;\mathbf{p}_{\perp})
= 2 i  e^{-p_{\perp}^2 \ell^2 }  \sum_{n=0}^{\infty} \frac{(-1)^n 
D_n(\omega,p^3;\mathbf{p}_{\perp})}{\omega^2-2n|eB| -(p^3)^2-m^2},
\label{prop-S2-momentum}
\end{equation}
where 
\begin{eqnarray}
D_n(\omega,p^3;\mathbf{p}_{\perp}) &=& \left(\gamma^0 \omega -\gamma^3 p^3 +m\right)
\left[ L_{n}\left(2p_{\perp}^2 \ell^2\right) {\cal P}_{+} 
-  L_{n-1}\left(2p_{\perp}^2 \ell^2\right){\cal P}_{-}\right] \nonumber\\
&& +2(\bm{\gamma}_{\perp}\cdot\mathbf{p}_\perp) L^{1}_{n-1}\left(2p_{\perp}^2 \ell^2\right).
\label{prop-S2-D-n-momentum}
\end{eqnarray}
Taking into account the earlier comment that the perpendicular momenta of charged 
particles are not conserved quantum numbers, this representation may appear surprising. 
However, one should keep in mind that the result in Eq.~(\ref{prop-S2-momentum}) is not 
a usual momentum representation of the propagator, but the Fourier transform of its 
translationary invariant part only. 

In some applications, it is convenient to make use of the so-called proper-time representation
\cite{Schwinger:1951nm}, in which the sum over Landau levels is traded for a proper-time
integration. This is easily derived from (\ref{prop-S2-momentum}) by making the following 
substitution:
\begin{equation}
\frac{i}{\omega^2-2n|eB| -(p^3)^2-m^2+i 0}  =  
\int_{0}^{\infty} ds \, 
e^{i s \left[\omega^2-2n|eB| -(p^3)^2-m^2+i 0\right]}.
\end{equation}
Then, the sum over Landau levels can be easily performed with the help of the 
summation formula for Laguerre polynomials \cite{1980tisp.book.....G},
\begin{equation}
 \sum_{n=0}^{\infty}L^{\alpha}_n(x)z^n= (1-z)^{-(\alpha+1)}\exp\left(\frac{xz}{z-1}\right).
 \label{LaguerreSum}
\end{equation}
The final expression for the propagator in the proper-time representation reads
\begin{eqnarray}
\tilde{S}(\omega,p^3;\mathbf{p}_{\perp})
&=&
 \int_{0}^{\infty} ds e^{i s\left[\omega^2-m^2-(p^3)^2\right]-i(p_\perp^2\ell^2) \tan(s|eB|)}
 \Big[\gamma^0 \omega -(\bm{\gamma}\cdot \mathbf{p}) +m
 \nonumber\\
&& 
+ (p^1\gamma^2-p^2\gamma^1) \tan(s eB) \Big]
 \left[1-  \gamma^1\gamma^2 \tan(seB)\right], 
\label{proper-time3+1}
\end{eqnarray}
where $(\bm{\gamma}\cdot \mathbf{p})\equiv (\bm{\gamma}_{\perp}\cdot \mathbf{p}_\perp)+\gamma^3 p^3$.

Using the same method, one can also derive the Dirac fermion propagator in a magnetic 
field in $2+1$ dimensions. It has the same structure as the propagator in Eqs.~(\ref{prop-S2}),
(\ref{schwinger-phase}), (\ref{prop-S2-trans}), and (\ref{prop-S2-F-n}), but with $p^3 =0$, i.e.,
\begin{eqnarray}
S_{2+1}(\omega;\mathbf{r},\mathbf{r}^\prime)
= e^{i\Phi(\mathbf{r},\mathbf{r}^{\prime})}
\tilde{S}_{2+1}(\omega;\mathbf{r}-\mathbf{r}^\prime) ,
\label{prop-S2+1}
\end{eqnarray}
where
\begin{eqnarray}
\tilde{S}_{2+1}(\omega;\mathbf{r}-\mathbf{r}^\prime)
&=& i \frac{e^{-\xi/2}}{2\pi\ell^2} \sum_{n=0}^{\infty} \Bigg[
\frac{\gamma^0 \omega +m}{\omega^2 -2n|eB| -m^2}
\left[ L_{n}(\xi) {\cal P}_{-}+ L_{n-1}(\xi){\cal P}_{+}\right] \nonumber\\
&&-\frac{i}{\ell^2} \frac{ \bm{\gamma}\cdot(\mathbf{r}-\mathbf{r}^{\prime})}
{\omega^2-2n|eB|-m^2} L_{n-1}^1(\xi)\Bigg] .
\label{prop-S2+1-trans}
\end{eqnarray}
The Fourier transform of the translationally invariant part is
\begin{eqnarray}
\tilde{S}_{2+1}(\omega;\mathbf{p})
&=& 2 i  e^{-p^2 \ell^2 }  \sum_{n=0}^{\infty} (-1)^n \Bigg[
\frac{\left(\gamma^0 \omega +m\right)\left[ L_{n}\left(2p^2 \ell^2\right) {\cal P}_{+} 
-  L_{n-1}\left(2p^2 \ell^2\right){\cal P}_{-}\right]}{\omega^2-2n|eB|-m^2}
\nonumber\\
&&+2\frac{(\bm{\gamma}\cdot\mathbf{p})}{\omega^2-2n|eB|-m^2}
L^{1}_{n-1}\left(2p^2 \ell^2\right)\Bigg].
\label{prop-S2+1-momentum}
\end{eqnarray}
Finally, the proper-time representation reads
\begin{eqnarray}
\tilde{S}_{2+1}(\omega;\mathbf{p})
&=&
 \int_{0}^{\infty} ds e^{i s\left[\omega^2-m^2\right]-i(p^2\ell^2) \tan(s|eB|)}
 \Big[\gamma^0 \omega -(\bm{\gamma}\cdot \mathbf{p}) +m
 \nonumber\\
&& 
+ (p^1\gamma^2-p^2\gamma^1) \tan(s eB) \Big]
 \left[1-  \gamma^1\gamma^2 \tan(seB)\right]. 
\label{proper-time-2+1}
\end{eqnarray}

\end{document}